\documentclass[journal,twoside]{IEEEtran}
\usepackage{cite}
\ifCLASSINFOpdf
   \usepackage[pdftex]{graphicx}
\else

\fi
\usepackage{graphicx}
\usepackage[cmex10]{amsmath}
\usepackage{algorithmic}
\usepackage{array}
\usepackage{mdwmath}
\usepackage{mdwtab}
\usepackage[tight,footnotesize]{subfigure}
\usepackage[font=footnotesize]{subfig}
\usepackage{fixltx2e}
\usepackage{url}
\usepackage{amsfonts}
\usepackage{epstopdf}
\usepackage{balance}
\usepackage{citesort}
\begin{document}

\title{Parametric Schemes for Prediction of Wideband MIMO Wireless Channels}
\author{Ramoni~Adeogun,~\IEEEmembership{Student~Member,~IEEE,}
        Paul~Teal,~\IEEEmembership{Senior~Member,~IEEE}
        and~Pawel~Dmochowski,~\IEEEmembership{Senior~Member,~IEEE}
\thanks{The authors are with the School of Engineering and Computer Science, Victoria University of Wellington, Wellington, New Zealand. e-mail: \{ramon.adeogun, pawel.dmochowski, paul.teal\}@ecs.vuw.ac.nz}
}

%



\maketitle

\begin{abstract}
Information on the future state of time varying frequency selective channels can significantly enhance the effectiveness of feedback in 
adaptive and limited feedback MIMO-OFDM systems. This paper investigates the parametric extrapolation of wideband MIMO channels using 
variations of  the double directional MIMO model. We propose three predictors which estimate parameters of the channel using 4D, 3D and 2D extensions 
of the ESPRIT algorithm and predict future states of the channel using the models. 
Furthermore, using the vector formulation of the Cramer Rao lower bound for functions of parameters, we derive a bound on the prediction error in 
wideband MIMO channels. Numerical simulations are used to evaluate the performance of the proposed algorithms under different channel and 
transmission conditions, and a comparison is made with the derived error bound.
\end{abstract}


\IEEEpeerreviewmaketitle

\section{Introduction}

\IEEEPARstart{T}{he} combination of MIMO transmission with Orthogonal Frequency Division Multiplexing (OFDM) \cite{li2006} is a spectrally 
efficient technique for achieving reliable high bit rate transmission over mobile wideband channels. It is employed in wireless standards such as 3GPP LTE and 
LTE Advanced \cite{Dahlman2008}, IEEE 802.16e (WiMAX) \cite{Wimax1,Gray06}. Recent capacity approaching MIMO-OFDM based transmission schemes, such as 
adaptive MIMO precoding \cite{Xia04}, adaptive coding and modulation, adaptive multiuser resource allocation and 
scheduling \cite{Kountouris2008,DBLP11} and various forms of codebook and non-codebook based limited feedback MIMO, require both at the transmitter and the receiver knowledge of the channel state information (CSI). In time division duplex (TDD)
systems, channel reciprocity is used to obtain CSI. In frequency division duplex (FDD) systems however, CSI
is estimated at the receiver and relayed in quantized form to the transmitter via a low rate feedback link.
In practical MIMO sytems, due to delays in estimation, processing and feedback, the CSI may become outdated before its actual use at the 
transmitter, resulting in significant performance degradation especially in high mobility environments. Prediction of the CSI has been recognised 
as an effective technique for mitigating this performance degradation \cite{Duel2000,Rhee20082095}.

The problem of channel prediction for single input single output (SISO) channels has been studied extensively. In \cite{Duel2000,Ekman2002,Oien2004}, 
the narrowband SISO channel is modelled as an autoregressive (AR) process and a linear minimum mean squared error (MMSE) predictor  
is used to extrapolate the channel states. These schemes consider the time-varying channel as a stochastic wide 
sense stationary process and use the temporal correlation for prediction without accounting for the physical scattering phenomena that cause the fading. 
Other researchers \cite{Anderson99,Wong08,Vaughan} have considered a ray based sum of sinusoids model where the fading channel is modelled as a sum of a finite number of plane waves. 
Extensions of these algorithms to wideband SISO channels have also been studied \cite{Liu2006,Yang2001}. Analytical and simulation results on SISO 
prediction have proven that with dense scattering, SISO channels can only be predicted over a very short distance (on the order of tenths of a 
wavelength) depending on the environment and propagation scenarios. The bound on SISO channel prediction error \cite{paul2001} indicates that
CSI is required over several wavelengths in order to accurately predict the channel. It has also been shown that prediction beyond a wavelength 
is not realistic, particularly in practical cases where the stationarity assumption does not hold for a time comparable to the duration of the observation.

The prediction of multi-antenna channels was first investigated in \cite{Arredondo02} through an evaluation of downlink beamforming
with channel prediction. Improved MISO channel prediction was shown, as more structure of the wavefield is revealed through multiple sampling. 
Bounds on the prediction error of MIMO channels \cite{Svantesson2006} and MIMO-OFDM channels \cite{Larsen2008} indicate that better prediction can be obtained by utilizing
the channel spatial structure. The authors illustrated this using  AR modelling for the prediction of beamspace
transformed CSI, and argue that the transformation reduces the effective number of rays present
in the channel, ultimately resulting  in longer prediction. A similar approach based on ray cancelling was presented in \cite{kenta08}.

MIMO prediction schemes can be broadly classified into codebook based precoder prediction and non-codebook based CSI prediction. 
The codebook based schemes \cite{ChangCodebook11,Inoue2009} predict the precoder for the next transmission frame using 
linear prediction. 
Others adopt Givens rotations to transform the precoding matrix and perform prediction of the Givens 
parameter \cite{Godana2011}. These schemes are limited to one step prediction and the channel model is often assumed to be
independent and identically distributed (i.i.d). Non-codebook based schemes predict the actual CSI using either autoregressive 
modelling or parametric model based SISO approaches \cite{Vanderpypen}. These methods do not utilize the additional spatial 
information that is revealed by the presence of multiple sensors. 


Motivated by the results in \cite{Larsen2008,Larsen2009}, where it was shown using error bounds that schemes which incorporate both temporal 
and spatial channel information offer significant improvement in prediction performance, we make the following contributions in this paper.
\begin{itemize}
\item Using a double directional spatial channel model, we derive three formulations of prediction models by progressively removing 
the dependence on array geometry. This allows the investigation of the effects of transmit and/or receive spatial dimensions on prediction 
performance, and the development of schemes applicable to systems with different antenna geometries.
\item We propose prediction schemes based on the above models and original adaptations of multidimensional ESPRIT (estimation of signal parameters 
via rotational invariance techniques) \cite{Roy2}. We show that this approach allows for the resolution of many more paths than the number of antennas at both ends of the link.
\item Using the vector formulation of the Cramer Rao bound for functions of parameters, we derive an expression for the bound on the 
prediction error. Although similar analyses have been presented in \cite{Larsen2008,Larsen2009}, our formulations are simpler and easier to interpret.
\end{itemize}
The remainder of the paper is organized as follows. Section~\ref{sec:channel} presents different formulations of the double directional channel model 
for wideband MIMO systems. In Section~\ref{sec:trans}, we describe the data transformation and preprocessing required for the implementation of 
the proposed schemes. The channel predictors are presented in Section~\ref{sec:algo}, followed by a derivation of the bound on prediction error 
in Section~\ref{sec:bound}. Performance evaluation results and discussion are given in Section~\ref{sec:simu}. Finally, we draw conclusions in Section~\ref{sec:conc}.

\section{Channel Models}\label{sec:channel}

We consider several formulations of a ray-based wideband spatial MIMO channel model for the development of the prediction schemes in this paper. 
The formulations are  extensions of the continuous time impulse response of doubly selective SISO fading channels, defined as
\begin{equation}
 \label{eq1}
 h(t;\tau)=\sum_{p=1}^{P}\alpha_{p}(t)\delta(\tau-\tau_p(t))
\end{equation}
where $t$ and $\tau$ are the time and delay variables respectively, $P$ is the number of paths, and $\alpha_p(t)$ and $\tau_p(t)$ are the time-varying complex attenuation and
delay of the $p$th path, respectively.  We assume that the scattering sources are in the far field of both the transmit and receive antenna arrays
such that the propagating waves can be modelled as plane waves. The complex attenuation of the $p$th path can thus be defined as
\begin{equation}
 \label{eq2}
 \alpha_p(t)=\sum_{r=1}^{R_p}\beta_{r,p}\exp(j\nu_{r,p}t)
\end{equation}
where $R_p$ is the number of rays in the $p$th path, $j=\sqrt{-1}$, $\beta_{r,p}$ and $\nu_{r,p}$ are the complex amplitude and Doppler frequency of 
the $r$th ray in the $p$th path, respectively. 
The model in \eqref{eq2} can be extended to a MIMO channel with $M$ transmit and $N$ receive antennas via the introduction of transmit and receive 
array structures, giving
\begin{equation}
 \label{eq3}
 \mathbf{H}_p(t)=\sum_{r=1}^{R_p}\beta_{r,p}\mathbf{a}_{\mathrm{r}}(\theta_{r,p})\mathbf{a}^T_{\mathrm{t}}(\phi_{r,p})\exp(j\nu_{r,p}t)
\end{equation}
where $[\cdot]^T$ denotes the non-conjugate transpose of the associated matrix, $\mathbf{a}_{\mathrm{r}}(\theta_{r,p})$ and 
$\mathbf{a}_{\mathrm{t}}(\phi_{r,p})$ are the receive and transmit array response vectors and $\theta_{r,p}$ and $\phi_{r,p}$  are the directions of arrival and directions of departure, respectively. 
Summing \eqref{eq3} over the clusters and taking the Fourier transform in the delay domain, we obtain the frequency response of the MIMO channel  as
\begin{align}
 \label{eq5}
 \mathbf{H}(t,f)&=\sum_{p=1}^{P}\mathbf{H}_p(t)\exp(-j2\pi f \tau_p)\nonumber\\
 &=\sum_{p=1}^{P}\sum_{r=1}^{R_p}\beta_{r,p}\mathbf{a}_{\mathrm{r}}(\theta_{r,p})\mathbf{a}^T_{\mathrm{t}}(\phi_{r,p})\exp(j\nu_{r,p}t-j2\pi f \tau_p)
\end{align}
where $f$ is the frequency variable. Assuming symbol duration $\Delta t$ and subcarrier spacing $\Delta f$, the sampled frequency response is given by
\begin{equation}
 \label{eq6}
 \mathbf{H}(q,k)=\sum_{p=1}^{P}\sum_{r=1}^{R_p}\beta_{r,p}\mathbf{a}_{\mathrm{r}}(\theta_{r,p})\mathbf{a}^T_{\mathrm{t}}(\phi_{r,p})\exp(jq\gamma_{r,p}-jk\eta_p)
\end{equation}
where $q=0,\cdots,Q-1$ and $k=0,\cdots, K-1$ are the time and subcarrier indices, respectively. $\gamma_{r,p}=\nu_{r,p}\Delta t$ and $\eta_p = 2\pi\Delta f \tau_p$ are the normalized radian Doppler frequency and delay, respectively. Combining indices in \eqref{eq6}, we obtain
\begin{equation}
 \label{eq7}
 \mathbf{H}(q,k)=\sum_{z=1}^{Z}\beta_{z}\mathbf{a}_r(\theta_z)\mathbf{a}^T_t(\phi_z)\exp(jq\gamma_{z}-jk\eta_z)
\end{equation}
where $Z=\sum_{p=1}^{P} R_p$ is the total number of propagating rays. Each ray is characterized by the parameter 
set $\{\beta_z,\theta_z,\phi_z,\gamma_z,\eta_z\}$. We assume that no two rays share a common parameter set, but different rays may have 
one or more equal parameters. Note that in practical scenarios, \eqref{eq7} has a finite support in the transmit angular, Doppler, delay and 
receive angular domains since the multipath parameters are bounded. 

We now describe different formulations of the model in \eqref{eq7}, where we progressively remove restrictions on the array structure. 
\subsection{DOD/DOA Model}
The first model is based on the assumption that the array response vectors $\mathbf{a}_{\mathrm{r}}$ and $\mathbf{a}_{\mathrm{t}}$ are explicit 
functions of the directions of arrival (DOA) and directions of departures (DOD) \cite{Larsen2009} as shown in \eqref{eq7}. Note that this model is 
valid for any array geometry. We will consider systems with uniform linear arrays (ULA) at both ends of the link. The receive steering vector 
for an $N$ element array is thus
\begin{equation}
 \label{eq8}
 \mathbf{a}_{\mathrm{r}}(\mu^{\mathrm{r}}_{z})=[1,\,\exp(j\mu^{\mathrm{r}}_{z}),\,\cdots,\,\exp(j(N-1)\mu^{\mathrm{r}}_{z})]^T
\end{equation}
where 
\begin{equation}
\label{eq8a}
 \mu^{\mathrm{r}}_{z}=2\pi d_{\mathrm{r}}\sin(\theta_{z}),
\end{equation}
$d_{\mathrm{r}}$ is the receive array element spacing. The $M\times 1$ transmit array steering vector is defined analogously, with
\begin{equation}
\label{eq8b}
 \mu^{\mathrm{t}}_{z}=2\pi d_{\mathrm{t}}\sin(\phi_{z})
\end{equation}
Here, $d_{\mathrm{t}}$  is the transmit array element spacing. Henceforth, the parameter set for the models will include the spatial 
frequencies $\mu^{\mathrm{r}}_{z}$ and $\mu^{\mathrm{t}}_{z}$ rather than the actual directions $\theta_z$ and $\phi_z$, since the latter can be
trivially obtained from $\mu^{\mathrm{r}}_{z}$, $\mu^{\mathrm{t}}_{z}$ using \eqref{eq8a} and \eqref{eq8b}.
\subsection{Transmit Spatial Signature Model (TSSM)}
The DOD/DOA model depends on the specific array configurations and the angles of arrival and departure. Since the transmitter is often stationary 
in mobile wireless systems, estimation of the angles of departure may be difficult and possibly not required for accurate prediction of the channel. 
We therefore replace the product of the complex amplitude and the transmit array steering vector for each path in \eqref{eq7} by an unstructured 
transmit spatial signature (TSS)\footnote{A similar model termed vector spatial signature (VSS) was used in \cite{Larsen2009}.} vector, $\mathbf{s}$. 
The model in \eqref{eq7} can now be expressed as
\begin{equation}
 \label{eq9}
 \mathbf{H}(q,k)=\sum_{z=1}^{Z}\mathbf{a}_{\mathrm{r}}(\mu^{\mathrm{r}}_{z})\mathbf{s}^T_z\exp(jq\gamma_{z}-jk\eta_z)
\end{equation}
where $\mathbf{s}_z\in\mathbb{C}^{M\times 1}$ is the TSS for the $z$th propagating wave.
\subsection{Matrix Spatial Signature Model (MSSM) }
Similar to the TSS model, the MSSM \cite{Larsen2009} replaces the product of the array steering vectors by an $N\times M$ unstructured matrix spatial signature, $\mathbf{S}$, giving
\begin{equation}
 \label{eq10}
 \mathbf{H}(q,k)=\sum_{z=1}^{Z}\mathbf{S}_z\exp(jq\gamma_{z}-jk\eta_z)
\end{equation}
Note that although the three models described in this section are derived from the double directional MIMO model,
they differ in the parametrization and number of parameters required. A summary of the number of parameters required for each model and dependence on number of antenna elements is shown in Table~\ref{tab:tab1}. The parameters
of the channel are assumed quasi-stationary over the spatial distance for which channel observation/measurement and prediction is made.
This assumption has been shown in the industry standard 3GPP/WINNER II SCM model \cite[p.~55]{WINNER2} to be valid for mobile movements
up to $50\lambda$. We also assume that $Q$ temporal samples and $K$ frequency samples of the channel frequency response matrix are available
by transmitting known training sequences or from other channel estimation approaches. 

In practice, the estimated or measured channel will be imperfect due to the effects of noise and interference. The estimated CSI matrix at 
time instant $q$ for the $k$th subcarrier is therefore defined as\footnote{Henceforth, we use $\hat{\,}$ to denote estimates corrupted by noise.}
\begin{equation}
\label{eq:eq5}
\mathbf{\hat{H}}(k,q)=\mathbf{H}(k,q)+\mathbf{N}(k,q)
\end{equation}
where $\mathbf{N}(k,q)\in \mathbb{C}^{N\times M}$ is a matrix of complex Gaussian random variables that accounts for channel estimation errors.

\begin{table}[!t]
\renewcommand{\arraystretch}{1.2}
\caption{MIMO Model Parametrization and Dependence on Number of Antennas}
\label{tab:tab1}
\centering
\begin{tabular}{|c||c|c|c|}
\hline
Model & Structural Param. & Amp. & Real Param.\\
\hline
DOD/DOA & $\{\mu^{\mathrm{r}}_{z},\mu^{\mathrm{t}}_{z},\gamma_z,\eta_z\}_{z=1}^Z$ & $\{\mathfrak{R}(\beta_z),\mathfrak{I}(\beta_z)\}_{z=1}^Z$ & $6Z$\\
\hline
TSSM & $\{\mu^{\mathrm{r}}_{z},\gamma_z,\eta_z\}_{z=1}^Z$ & $\{\mathfrak{R}(\mathbf{s}_z),\mathfrak{I}(\mathbf{s}_z)\}_{z=1}^Z$ & $Z(2M+3)$\\
\hline
MSSM & $\{\gamma_z,\eta_z\}_{z=1}^Z$ & $\{\mathfrak{R}(\mathbf{S}_z),\mathfrak{I}(\mathbf{S}_z)\}_{z=1}^Z$ & $2Z(NM+1)$\\
\hline
\end{tabular}
\end{table}
\section{Data Transformation}\label{sec:trans}
Having described the channel model variations for the development of the prediction schemes, we now present the data preprocessing necessary for 
extraction of the parameters from available channel observations.
\subsection{DOD/DOA Transformation}
As shown in \eqref{eq7}, the DOD/DOA model is characterized by $4Z$ structural parameters $\{\mu^{\mathrm{r}}_{z},\mu^{\mathrm{t}}_{z},\gamma_z,\eta_z\}_{z=1}^Z$ and $Z$ 
complex amplitudes. Extraction of these parameters from the channel observations requires a four dimensional array data structure. 
Since we consider MIMO systems with a 1-D antenna array (i.e., a ULA) at both ends of the link, it is necessary to convert the channel matrices 
into a form that allows 4D parameter estimation to be performed. Let $\mathbf{h}(q,k)=\operatorname{vec}[\mathbf{H}(q,k)]\in\mathbb{C}^{NM\times 1}$ 
be a vector obtained by stacking the columns of $\mathbf{H}(q,k)$. Using \eqref{eq7} and the properties of the Kronecker product, it can be shown 
that
\begin{equation}
\label{eq11}
\mathbf{h}(q,k)=\sum_{z=1}^{Z}\beta_{z}(\mathbf{a}_{\mathrm{r}}(\mu^{\mathrm{r}}_{z})\otimes\mathbf{a}_{\mathrm{t}}(\mu^{\mathrm{t}}_{z})\exp(jq\gamma_{z}-jk\eta_z)
\end{equation}
where $\otimes$ denotes the Kronecker product. Note that the transformation in \eqref{eq11} combines the receive and transmit spatial dimension of 
the channel. In order to introduce the temporal dimension, we define
\begin{equation}
\label{eq12}
\mathbf{D}(k)=\left[\mathbf{h}(1,k)\quad\mathbf{h}(2,k)\quad\cdots\quad\mathbf{h}(Q,k)\right]
\end{equation}
and form a Hankel matrix by sliding an $NM\times R$ rectangular window through \eqref{eq12} to obtain
\begin{equation}
\label{eq13}
\mathbf{D}_k=\begin{bmatrix}
\mathbf{h}(1,k) & \mathbf{h}(2,k) & \cdots & \mathbf{h}(R,k)\\
\mathbf{h}(2,k) & \mathbf{h}(3,k) & \cdots & \mathbf{h}(R+1,k)\\
\vdots & \vdots & \ddots & \vdots\\
\mathbf{h}(S,k) & \mathbf{h}(S+1,k) & \cdots & \mathbf{h}(Q,k)\\
\end{bmatrix}
\end{equation}
where $S=Q-R+1$. The frequency dimension of the channel is similarly introduced by forming a block Hankel matrix from $K$ such matrices to obtain
\begin{equation}
\label{eq14}
\mathbf{X}_{\mathrm{d}}=\begin{bmatrix}
\mathbf{D}_1 & \mathbf{D}_2 & \cdots & \mathbf{D}_T\\
\mathbf{D}_2 & \mathbf{D}_3 & \cdots & \mathbf{D}_{(T+1)}\\
\vdots & \vdots & \ddots & \vdots\\
\mathbf{D}_U & \mathbf{D}_{(U+1)} & \cdots & \mathbf{D}_K
\end{bmatrix}
\end{equation}
where $U$ and $T$ are the Hankel matrix size parameters with $U=K-T+1$. The values of $S$, $R$,  $T$, and $U$ are selected such that $NMSU\geq Z+1$. 
There is, however, a compromise in selecting these: large values of $S$ and $U$ increases the number of rows in $X_{\mathrm{d}}$ and 
hence the number of sources that can be resolved, but this results in small values of $R$ and $T$ which affects the accuracy of covariance estimates. 
Using the model in \eqref{eq11} and the transformations in \eqref{eq13} and \eqref{eq14}, the data in the columns of $\mathbf{X}_{\mathrm{d}}$ can be modelled as
\begin{equation}
\label{eq15}
\mathbf{x}_{\mathrm{d}}(i) = \sum_{z=1}^{Z}\beta_z\mathbf{a}(\mu^{\mathrm{r}}_{z},\mu^{\mathrm{t}}_{z},\gamma_z,\eta_z)\exp(-j(i-1)\eta_z)
\end{equation}
where $\mathbf{a}(\mu^{\mathrm{r}}_{z},\mu^{\mathrm{t}}_{z},\gamma_z,\eta_z)=(\mathbf{a}_r(\mu^{\mathrm{r}}_{z})\otimes\mathbf{a}_t(\mu^{\mathrm{t}}_{z})\otimes\mathbf{a}_d(\gamma_z)\otimes\mathbf{a}_{\tau}(\eta_z))$ with
\begin{align}
\label{eq16}
\mathbf{a}_{\mathrm{d}}(\gamma_z)&=[1 \quad\exp(j\gamma_z)\quad\cdots\quad\exp(j(R-1)\gamma_z)]^T\nonumber\\
\mathbf{a}_{\tau}(\eta_z)&=[1 \quad\exp(-j\eta_z)\quad\cdots\quad\exp(-j(U-1)\eta_z)]^T
\end{align}
Defining $\alpha_z(i) = \beta_z\exp(-j(i-1)\eta_z)$, \eqref{eq15} can be expressed as
\begin{align}
\label{eq17}
\mathbf{x}_{\mathrm{d}}(i) &= \sum_{z=1}^{Z} \alpha_z(i)\mathbf{a}(\mu^{\mathrm{r}}_{z},\mu^{\mathrm{t}}_{z},\gamma_z,\eta_z)\nonumber\\
&=\mathbf{A}(\boldsymbol{\mu}^{\mathrm{r}},\boldsymbol{\mu}^{\mathrm{t}},\boldsymbol{\gamma},\boldsymbol{\eta})\boldsymbol{\alpha}(i)
\end{align}
where $\boldsymbol{\alpha}(i)=[\alpha_1\quad\cdots\quad\alpha_Z]\in\mathbb{C}^{Z\times 1}$ and
$\mathbf{A}=[\mathbf{a}(\mu^{\mathrm{r}}_{1},\mu^{\mathrm{t}}_{1},\gamma_1,\eta_1)\quad\cdots\quad\mathbf{a}(\mu^{\mathrm{r}}_{Z},\mu^{\mathrm{t}}_{Z}
,\gamma_Z,\eta_Z)]$ is a Vandermonde structured steering matrix, with $\boldsymbol{\mu}^{\mathrm{r}},\boldsymbol{\mu}^{\mathrm{t}},\boldsymbol{\gamma},\boldsymbol{\eta}$
defined as $Z\times 1$ vectors of their respective parameters. 
Clearly, \eqref{eq15} corresponds to a four dimensional array data model obtained by combining the transmit spatial, temporal, frequency and receive spatial dimensions of the wideband MIMO channel. A summary of the dimensions and corresponding parameters is shown in Table~\ref{tab:tab2}.
\begin{table}[!t]
\renewcommand{\arraystretch}{1.5}
\caption{Wideband MIMO Data Domain and Parameters}
\label{tab:tab2}
\centering
\begin{tabular}{|c||c|c|c|c|}
\hline
Model & \multicolumn{4}{|c|}{Data Domain}\\
\hline
 & Receive  & Transmit & Temporal & Frequency\\
 \hline
DOA/DOD & AOA & AOD & Doppler Shift & Delay \\
\hline
TSSM & AOA & - & Doppler shift& Delay\\
\hline
MSSM & - & - & Doppler shift & Delay\\
\hline
\end{tabular}
\end{table}
\subsection{TSS Transformation}
As shown in \eqref{eq9} and Table~\ref{tab:tab1}, parametrizing the TSSM requires $3Z$ structural parameters $\{\mu^{\mathrm{r}}_{z},\gamma_z,\eta_z\}_{z=1}^Z$. 
We will here derive a data structure that allows joint extraction of these parameters from \eqref{eq9}. Using the $Q$ temporal samples, we start by 
forming a block Hankel matrix for each frequency sample,
\begin{equation}
\label{eq18}
\mathbf{B}_k = \begin{bmatrix}
\mathbf{H}(1,k) & \mathbf{H}(2,k) & \cdots & \mathbf{H}(R,k)\\
\mathbf{H}(2,k) & \mathbf{H}(3,k) & \cdots & \mathbf{H}(R+1,k)\\
\vdots & \vdots & \ddots & \vdots\\
\mathbf{H}(S,k) & \mathbf{H}(S+1,k) & \cdots & \mathbf{H}(Q,k)\\
\end{bmatrix}
\end{equation}
Note that \eqref{eq18} combines the receive spatial and temporal property of the channel into one dimension corresponding to the columns 
of $\mathbf{B}_k$. In order to include the frequency dimension of the wideband channel, we form another block Hankel matrix from \eqref{eq18} as
\begin{equation}
\label{eq19}
\mathbf{X}_{\mathrm{t}}=\begin{bmatrix}
\mathbf{B}_1 & \mathbf{B}_2 & \cdots & \mathbf{B}_{T}\\
\mathbf{B}_2 & \mathbf{B}_3 & \cdots & \mathbf{B}_{T+1}\\
\vdots & \vdots & \ddots & \vdots\\
\mathbf{B}_{U} & \mathbf{B}_{U+1} & \cdots & \mathbf{B}_{K}\\
\end{bmatrix}
\end{equation}
Based on the transformations in \eqref{eq18} and \eqref{eq19}, the data in the columns of $\mathbf{X}_{\mathrm{t}}$ is given by
\begin{equation}
\label{eq20}
\mathbf{x}_{\mathrm{t}}(i)=\mathbf{A}(\boldsymbol{\mu}^{\mathrm{r}},\boldsymbol{\gamma},\boldsymbol{\eta})\boldsymbol{\alpha}(i)
\end{equation}
where $\mathbf{A}(\boldsymbol{\mu}^{\mathrm{r}},\boldsymbol{\gamma},\boldsymbol{\eta})$ is defined analogously to \eqref{eq17}.\footnote{For simplicity of notation, we use a common variable 
for the array steering matrix in all three models. The precise definition is obvious from the context.}
\subsection{MSS Transformation}
The MSSM is parametrized by $2Z$ structural parameters and $NMZ$ complex amplitude parameters. Extraction of these parameters from the 
channel requires a two-dimensional datum. Similar to \eqref{eq12}, we form a Hankel matrix
\begin{equation}
\label{eq23}
\mathbf{C}_k=\begin{bmatrix}
\mathbf{h}^T(1,k) & \mathbf{h}^T(2,k) & \cdots & \mathbf{h}^T(R,k)\\
\mathbf{h}^T(2,k) & \mathbf{h}^T(3,k) & \cdots & \mathbf{h}^T(R+1,k)\\
\vdots & \vdots & \ddots & \vdots\\
\mathbf{h}^T(S,k) & \mathbf{h}^T(S+1,k) & \cdots & \mathbf{h}^T(Q,k)\\
\end{bmatrix}
\end{equation}
Note that the columns of $\mathbf{C}_k$ correspond to $S$ temporal measurements of the channel and can be modelled as
\begin{equation}
\label{eq24}
\mathbf{c}_k(i)=\sum_{z=1}^{Z}\alpha_z(i)\mathbf{a}_d(\gamma_z)
\end{equation}
where $\mathbf{a}_d(\gamma_z)$ is an $S$ dimensional vector defined in \eqref{eq16}. 
The data in \eqref{eq24} provides information about the Doppler shifts of the channel. The frequency structure of the channel can be included by 
forming a block Hankel matrix $\mathbf{X}_{\mathrm{m}}$ analogous to $\mathbf{X}_{\mathrm{d}}$ with $\mathbf{D}_k$ replaced with $\mathbf{C}_k$. 
The columns of $\mathbf{X}_{\mathrm{m}}$ can be shown using \eqref{eq24} to be
\begin{align}
\label{eq26}
\mathbf{x}_{\mathrm{m}}(i) &= \sum_{z=1}^{Z}\alpha_z(i)(\mathbf{a}_d(\gamma_z)\otimes \mathbf{a}_{\tau}(\eta_z))\nonumber\\
&=\mathbf{A}(\boldsymbol{\gamma},\boldsymbol{\eta})\boldsymbol{\alpha}(i)
\end{align}
As shown in \eqref{eq26}, $\mathbf{X}_{\mathrm{m}}$ corresponds to a two-dimensional datum obtained by combining the temporal and frequency structure of
the channel. The Doppler shifts and delays of arrival can therefore be extracted jointly using appropriate parameter estimation algorithms.
\section{Prediction Algorithms}\label{sec:algo}
We now propose prediction algorithms using the models developed in Section \ref{sec:channel} and the transformed data derived in Section \ref{sec:trans}.
We will henceforth refer to the algorithms as WIMEMCHAP: WIdeband Multidimensional Esprit based Mimo CHAnnel Predictor, and use the acronyms for the models
as prefixes to distinguish the schemes. For example, the algorithm based on TSSM will be called TSSM-WIMEMCHAP and so on. Note that although the
algorithms are based on the same idea of parametric modelling, they differ in the model, dimension of parameter estimation and number of amplitude and structural
parameters to be estimated.
\subsection{DOD/DOA-WIMEMCHAP}
Consider the transformed data model in \eqref{eq15} and \eqref{eq17}. Since the array steering matrix $\mathbf{A}$ is equivalent to a product of
four Vandermonde matrices, the invariance structure in $\mathbf{A}$ can be utilized to estimate the parameters of the channel. Motivated by the accuracy
and computational efficiency of the ESPRIT algorithm \cite{Roy2}, we propose an adaptation of multidimensional extension of ESPRIT to jointly 
extract the parameter sets $\{\mu^{\mathrm{r}}_{z},\mu^{\mathrm{t}}_{z},\gamma_z,\eta_z\}_{z=1}^Z$ and apply the parameter estimates to extrapolate 
the channel. The prediction algorithm can be divided into the following stages:
\begin{itemize}
 \item covariance matrix estimation and subspace separation,
 \item number of paths estimation,
 \item joint parameter estimation,
 \item channel extrapolation.
\end{itemize}
We will now describe the different stages of the algorithm.
\subsubsection{Covariance Matrix Estimation}
In the presence of estimation or measurement noise, the model in \eqref{eq17} becomes
\begin{equation}
\label{eq27}
\hat{\mathbf{x}}_{\mathrm{d}}(i)=\mathbf{A}(\boldsymbol{\mu}^{\mathrm{r}},\boldsymbol{\mu}^{\mathrm{t}},\boldsymbol{\gamma},\boldsymbol{\eta})\boldsymbol{\alpha}(i)
+\mathbf{n}(i)
\end{equation}
where $\mathbf{n}(i)$ models the effects of $\mathbf{N}$ in \eqref{eq:eq5}. The covariance matrix (containing the spatial, temporal and frequency correlations) can be estimated from \eqref{eq27} as
\begin{align}
 \label{eq28}
 \hat{\mathbf{C}}_{\mathrm{d}}&=\frac{1}{RT}\sum_{i=1}^{RT}\hat{\mathbf{x}}_{d}(i)\hat{\mathbf{x}}_{d}(i)^H\nonumber\\
 &=\mathbf{A}(\boldsymbol{\mu}^{\mathrm{r}},\boldsymbol{\mu}^{\mathrm{t}},\boldsymbol{\gamma},\boldsymbol{\eta})\mathbf{C}_{\alpha\alpha}
 \mathbf{A}(\boldsymbol{\mu}^{\mathrm{r}},\boldsymbol{\mu}^{\mathrm{t}},\boldsymbol{\gamma},\boldsymbol{\eta})^H+\sigma^2\mathbf{I}
\end{align}
where $(\cdot)^H$ denotes the Hermitian transpose and $\mathbf{C}_{\alpha\alpha}=\frac{1}{RT}\sum_{i=1}^{RT}\boldsymbol{\alpha}\boldsymbol{\alpha}^H$ is the
covariance matrix of the amplitude parameters. $\hat{\mathbf{C}}_{\mathrm{d}}$ can be expressed in terms of $\hat{\mathbf{X}}_{\mathrm{d}}$ in \eqref{eq14} as
\begin{equation}
 \label{eq29}
 \hat{\mathbf{C}}_{\mathrm{d}}=\frac{1}{RT}\hat{\mathbf{{X}}}_{\mathrm{d}}\hat{\mathbf{{X}}}_{\mathrm{d}}^H
\end{equation}
\subsubsection{Number of Paths Estimation and Subspace Decomposition}
Estimation of the number of paths $Z$ is typically a model order selection problem. The Minimum Description Length (MDL)  is often used for this
purpose due to its accuracy and consistency \cite{Wax85,Kay}. We utilize a modified version of the MDL referred to as the Minimum Mean Square Error (MMSE)-MDL
\cite{Huang_mmsemdl} defined as
\begin{equation}
\label{eq:eq30}
\hat{Z}=\operatorname{arg}\vspace{-2pt}\min_{z=1,\cdots,NMSU-1} RT\log(\lambda_z)+\frac{1}{2}(z^2+z)\log RT
\end{equation}
where $\lambda_z$ are the eigenvalues of $\hat{\mathbf{C}}_{\mathrm{d}}$. Once $\hat{Z}$ has been estimated, the eigenvalue decomposition of
${\mathbf{C}}_{\mathrm{d}}$ can be expressed as
\begin{align}
\label{eq:eqchap32}
\mathbf{C}_{\mathrm{d}}&=\left[\mathbf{E}_{\mathrm{s}}\quad \mathbf{E}_{\mathrm{n}}\right]\begin{bmatrix}
\boldsymbol{\Lambda}_{\mathrm{s}} & \quad\\ \quad & \boldsymbol{\Lambda}_{\mathrm{n}}
\end{bmatrix}\begin{bmatrix}
\mathbf{E}_{\mathrm{s}}^H\\ \mathbf{E}_{\mathrm{n}}^H
\end{bmatrix}\nonumber\\
&=\mathbf{E}_{\mathrm{s}}\boldsymbol{\Lambda}_{\mathrm{s}}\mathbf{E}_{\mathrm{s}}^H+\mathbf{E}_{\mathrm{n}}\boldsymbol{\Lambda}_{\mathrm{n}}\mathbf{E}_{\mathrm{n}}^H
\end{align}
where $\mathbf{E}_{\mathrm{s}}$ and $\boldsymbol{\Lambda}_{\mathrm{s}}$ are the signal subspace eigenvectors and the associated
eigenvalues, respectively. The noise subspace eigenvectors and eigenvalues are contained in $\mathbf{E}_{\mathrm{n}}$ 
and $\boldsymbol{\Lambda}_{\mathrm{n}}$, respectively.
\subsubsection{Parameter Estimation}
We now outline the process of obtaining the parameter sets $\{\mu^{\mathrm{r}}_{z},\mu^{\mathrm{t}}_{z},\gamma_z,\eta_z\}_{z=1}^{\hat{Z}}$ in Table~\ref{tab:tab1}.  
In order to explore the invariance structure \cite{Roy2} in the Vandermonde structured space-time-frequency manifold matrix, $\mathbf{A}$, we define the 
following selection matrices:
\begin{align}
\label{eq:eqPPE3}
\mathbf{S}_{1\theta}&=\begin{bmatrix} \mathbf{I}_{(N-1)} & \mathbf{0}_{(N-1)}\end{bmatrix} &\quad\quad \mathbf{S}_{\theta 1}&=\mathbf{I}_{M}\otimes\mathbf{I}_{S}\otimes\mathbf{I}_{U}\otimes \mathbf{S}_{1\theta}\nonumber\\
\mathbf{S}_{2\theta}&=\begin{bmatrix} \mathbf{0}_{(N-1)} & \mathbf{I}_{(N-1)} \end{bmatrix} &\quad\quad \mathbf{S}_{\theta 2}&=\mathbf{I}_{M}\otimes\mathbf{I}_{S}\otimes\mathbf{I}_{U}\otimes \mathbf{S}_{2\theta} \nonumber\\
\mathbf{S}_{1\phi}&=\begin{bmatrix} \mathbf{I}_{(M-1)} & \mathbf{0}_{(M-1)}\end{bmatrix}&\quad\quad \mathbf{S}_{\phi 1}&=\mathbf{I}_{S}\otimes\mathbf{I}_{U}\otimes \mathbf{S}_{1\varphi}\otimes\mathbf{I}_N \nonumber\\
\mathbf{S}_{2\phi}&=\begin{bmatrix} \mathbf{0}_{(M-1)} & \mathbf{I}_{(M-1)} \end{bmatrix}&\quad\quad  \mathbf{S}_{\phi 2}&=\mathbf{I}_{S}\otimes\mathbf{I}_{U}\otimes J_{2\phi}\otimes\mathbf{I}_N\nonumber\\
\mathbf{S}_{1\nu}&=\begin{bmatrix} \mathbf{I}_{(S-1)} & \mathbf{0}_{(S-1)}\end{bmatrix} &\quad\quad \mathbf{S}_{\nu 1}&=\mathbf{I}_{U}\otimes \mathbf{S}_{1\nu}\otimes\mathbf{I}_N\otimes\mathbf{I}_M \nonumber\\
\mathbf{S}_{2\nu}&=\begin{bmatrix} \mathbf{0}_{(S-1)} & \mathbf{I}_{(S-1)} \end{bmatrix} &\quad\quad \mathbf{S}_{\phi 2}&=\mathbf{I}_{U}\otimes \mathbf{S}_{2\nu}\otimes\mathbf{I}_N\otimes\mathbf{I}_M\nonumber\\
\mathbf{S}_{1\tau}&=\begin{bmatrix} \mathbf{I}_{(U-1)} & \mathbf{0}_{(U-1)}\end{bmatrix} &\quad\quad \mathbf{S}_{\tau 1}&=\mathbf{S}_{1\tau}\otimes \mathbf{I}_N\otimes\mathbf{I}_M\otimes\mathbf{I}_{S} \nonumber\\
\mathbf{S}_{2\tau}&=\begin{bmatrix} \mathbf{0}_{(U-1)} & \mathbf{I}_{(U-1)} \end{bmatrix} &\quad\quad \mathbf{S}_{\tau 2}&=\mathbf{S}_{2\tau}\otimes \mathbf{I}_N\otimes\mathbf{I}_M\otimes\mathbf{I}_{S}
\end{align}
where $\mathbf{I}_F$ is an $F\times F$ identity matrix and $\mathbf{0}_F \in \mathbf{R}^F$ is an F-dimensional vector of zeros. Using the selection 
matrices in \eqref{eq:eqPPE3}, we define the following invariance equations
\begin{align}
\label{eq:eqPPE4}
 \mathbf{S}_{\theta 2}\mathbf{E}_s&=\mathbf{S}_{\theta 1}\mathbf{E}_s\boldsymbol{\Phi}_{\mathrm{r}}&\quad\quad
 \mathbf{S}_{\phi 2}\mathbf{E}_s&=\mathbf{S}_{\phi 1}\mathbf{E}_s\boldsymbol{\Phi}_{\mathrm{t}}\nonumber\\
 \mathbf{S}_{\nu 2}\mathbf{E}_s&=\mathbf{S}_{\nu 1}\mathbf{E}_s\boldsymbol{\Phi}_{\mathrm{d}}&\quad\quad
 \mathbf{S}_{\tau 2}\mathbf{E}_s&=\mathbf{S}_{\tau 1}\mathbf{E}_s\boldsymbol{\Phi}_{\mathrm{f}}
\end{align}
where  $\boldsymbol{\Phi}_{\mathrm{r}}$, $\boldsymbol{\Phi}_{\mathrm{t}}$, $\boldsymbol{\Phi}_{\mathrm{d}}$ and $\boldsymbol{\Phi}_{\mathrm{f}}$ are 
matrices, the eigenvalues of which contain information about the parameters: 
\begin{align}
\label{eq:eqPPE5}
\operatorname{eig}(\boldsymbol{\Phi}_{\mathrm{r}})&=\operatorname{diag}\left[\exp(j\mu^{\mathrm{r}}_1),\exp(j\mu^{\mathrm{r}}_2),\cdots,\exp(j\mu^{\mathrm{r}}_{\hat{Z}})\right]\nonumber\\
\operatorname{eig}(\boldsymbol{\Phi}_{\mathrm{t}})&=\operatorname{diag}\left[\exp(j\mu^{\mathrm{t}}_1),\exp(j\mu^{\mathrm{t}}_2),\cdots,\exp(j\mu^{\mathrm{t}}_{\hat{Z}})\right]\nonumber\\
\operatorname{eig}(\boldsymbol{\Phi}_{\mathrm{d}})&=\operatorname{diag}\left[\exp(j\gamma_1),\exp(j\gamma_2),\cdots,\exp(j\gamma_{\hat{Z}})\right]\nonumber\\
\operatorname{eig}(\boldsymbol{\Phi}_{\mathrm{f}})&=\operatorname{diag}\left[\exp(j\eta_1),\exp(j\eta_2),\cdots,\exp(j\eta_{\hat{Z}})\right]
\end{align}
where $\operatorname{eig}(\cdot)$ denotes the diagonal eigenvalue matrix of the associated matrix. We minimize the squared error of the equations 
in \eqref{eq:eqPPE4} to obtain
\begin{align}
 \label{eq:eqPPE6a}\boldsymbol{\Phi}_{\mathrm{r}}&=((\mathbf{S}_{\theta 2}\mathbf{E}_s)^H(\mathbf{S}_{\theta 2}\mathbf{E}_s))^{-1}(\mathbf{S}_{\theta 2}\mathbf{E}_s)^H(\mathbf{S}_{\theta 1}\mathbf{E}_s)\\
  \label{eq:eqPPE6b}\boldsymbol{\Phi}_{\mathrm{t}}&=((\mathbf{S}_{\phi 2}\mathbf{E}_s)^H(\mathbf{S}_{\phi 2}\mathbf{E}_s))^{-1}(\mathbf{S}_{\phi 2}\mathbf{E}_s)^H(\mathbf{S}_{\phi 1}\mathbf{E}_s)\\
  \label{eq:eqPPE6c}\boldsymbol{\Phi}_{\mathrm{d}}&=((\mathbf{S}_{\nu 2}\mathbf{E}_s)^H(\mathbf{S}_{\nu 2}\mathbf{E}_s))^{-1}(\mathbf{S}_{\nu 2}\mathbf{E}_s)^H(\mathbf{S}_{\nu 1}\mathbf{E}_s)\\
   \label{eq:eqPPE6d}\boldsymbol{\Phi}_{\mathrm{f}}&=((\mathbf{S}_{\tau 2}\mathbf{E}_s)^H(\mathbf{S}_{\tau 2}\mathbf{E}_s))^{-1}(\mathbf{S}_{\tau 2}\mathbf{E}_s)^H(\mathbf{S}_{\tau 1}\mathbf{E}_s)
\end{align}
Estimates of the AOAs, AODs, Doppler shifts and delays could be obtained directly from the solutions of \eqref{eq:eqPPE6a}--\eqref{eq:eqPPE6d} 
followed by an additional pairing stage. In order to achieve automatic pairing of the estimates, we utilize a scheme similar to the mean eigenvalue 
decomposition (MEVD) pairing scheme \cite{Kikuma}. Defining
\begin{equation}
\label{eq:eqPPE7}
\boldsymbol{\Upsilon}= \boldsymbol{\Phi}_{\mathrm{r}}+\boldsymbol{\Phi}_{\mathrm{t}}+\boldsymbol{\Phi}_{\mathrm{d}}+\boldsymbol{\Phi}_{\mathrm{f}}
\end{equation}
we perform eigenvalue decomposition of $\boldsymbol{\Upsilon}$ to obtain the common eigenvectors of the four matrices in the sum
\begin{equation}
\label{eq:eqPPE8}
\boldsymbol{\Upsilon}=\boldsymbol{\Sigma}\boldsymbol{\Lambda}\boldsymbol{\Sigma}^{-1}
\end{equation}
The diagonal eigenvalue matrices are then obtained using
\begin{align}
\label{eq:eqPPE9a}\boldsymbol{\Xi}_\theta&=\boldsymbol{\Sigma}^{-1}\boldsymbol{\Phi}_{\mathrm{r}}\boldsymbol{\Sigma}\\
\label{eq:eqPPE9b}\boldsymbol{\Xi}_\phi&=\boldsymbol{\Sigma}^{-1}\boldsymbol{\Phi}_{\mathrm{t}}\boldsymbol{\Sigma}\\
\label{eq:eqPPE9c}\boldsymbol{\Xi}_\nu&=\boldsymbol{\Sigma}^{-1}\boldsymbol{\Phi}_{\mathrm{d}}\boldsymbol{\Sigma}\\
\label{eq:eqPPE9d}\boldsymbol{\Xi}_\tau&=\boldsymbol{\Sigma}^{-1}\boldsymbol{\Phi}_{\mathrm{f}}\boldsymbol{\Sigma}
\end{align}
where $\boldsymbol{\Xi}_\theta=\operatorname{eig}(\boldsymbol{\Phi}_{\mathrm{r}})$, $\boldsymbol{\Xi}_\phi=\operatorname{eig}(\boldsymbol{\Phi}_{\mathrm{t}})$, 
$\boldsymbol{\Xi}_\nu=\operatorname{eig}(\boldsymbol{\Phi}_{\mathrm{d}})$ and $\boldsymbol{\Xi}_\tau=\operatorname{eig}(\boldsymbol{\Phi}_{\mathrm{f}})$. Finally, estimates of the 
parameters are evaluated from \eqref{eq:eqPPE5} as
\begin{align}
\label{eq:eqPPE10a}\hat{\boldsymbol{\mu}}^{\mathrm{r}}&=-\operatorname{arg}(\operatorname{diag}(\boldsymbol{\Xi}_\theta))\\
\label{eq:eqPPE10b}\hat{\boldsymbol{\gamma}}&=\operatorname{arg}(\operatorname{diag}(\boldsymbol{\Xi}_\nu))\\
\label{eq:eqPPE10c}\hat{\boldsymbol{\mu}}^{\mathrm{t}}&=-\operatorname{arg}(\operatorname{diag}(\boldsymbol{\Xi}_\phi))\\
\label{eq:eqPPE10d}\hat{\boldsymbol{\eta}}&=-\operatorname{arg}(\operatorname{diag}(\boldsymbol{\Xi}_\tau))
\end{align}
\subsubsection{Complex Amplitude Estimation}
We assume that the complex amplitude of each path is equal for all antenna pairs, which is reasonable considering the separation 
of gain $\beta_z$ from array dependent steering vectors $\mathbf{a}_{\mathrm{r}}$ and $\mathbf{a}_{\mathrm{t}}$ in \eqref{eq7}. The complex amplitudes can therefore be estimated via a
least square fit to the known channel. Using the Vandermonde structure of
$\mathbf{A}$ and \eqref{eq7}, we form the following equation for the first entry of $\hat{\mathbf{H}}$ in \eqref{eq:eq5} for the first subcarrier
\begin{equation}
 \label{CAE1}
 \begin{bmatrix}
  \hat{h}_{11}(1)\\ \hat{h}_{11}(2)\\ \vdots\\ \hat{h}_{11}(Q)
 \end{bmatrix}
 = \begin{bmatrix}
    1  & \cdots & 1\\
    e^{j\hat{\gamma}_1}  & \cdots & e^{j\hat{\gamma}_{\hat{Z}}}\\
    \vdots  & \ddots & \vdots\\
    e^{j(Q-1)\hat{\gamma}_1}  & \cdots & e^{j(Q-1)\hat{\gamma}_{\hat{Z}}}
   \end{bmatrix}
   \begin{bmatrix}
    \beta_1\\ \beta_2\\ \vdots\\ \beta_{\hat{Z}}
   \end{bmatrix}+\begin{bmatrix}
   n(1)\\ n(2)\\ \vdots\\ n(Q)
   \end{bmatrix}
\end{equation}
which can be written in matrix form as
\begin{equation}\label{CAE2}
 \hat{\mathbf{h}}_{11}=\hat{\mathbf{W}}\boldsymbol{\beta}+\mathbf{n}
\end{equation}
$\boldsymbol{\beta}$ can be obtained via a regularized LS solution of \eqref{CAE2} as
\begin{equation}
 \label{CAE3}
 \hat{\boldsymbol{\beta}}=(\hat{\mathbf{W}}^H\hat{\mathbf{W}}+\sigma\mathbf{I})\hat{\mathbf{W}}^H\hat{\mathbf{h}}_{11}
\end{equation}
where $\sigma$ is the regularization parameter choosen to minimize the effects of errors in $\hat{\mathbf{W}}$ on the estimation. For the rest of this paper
$\sigma$ is chosen emprically as $10^{-5}$. Note that although \eqref{CAE3} gives an estimate of the complex amplitudes, our preliminary simulations
show that improved estimates can be obtained by using more entries of $\hat{\mathbf{H}}$ in the estimation. We therefore generalize \eqref{CAE2} as
\begin{equation}\label{CAE4}
 \hat{\mathbf{h}}_{nm}=\hat{\mathbf{W}}_{nm}\boldsymbol{\beta}+\mathbf{n}\quad \forall n\in [1,N]\quad m\in[1,M]
\end{equation}
with $\hat{\mathbf{W}}_{nm}$ defined as
\begin{equation}
 \label{CAE5}
 \hat{\mathbf{W}}_{nm} = \mathbf{v}^r_{n}\diamond\hat{\mathbf{W}}\diamond\mathbf{v}^t_{m}
\end{equation}
where $\diamond$ denotes the Khatri-Rao product, and  $\mathbf{v}^r_{n}$ is defined as
\begin{equation}
 \label{CAE6}
 \mathbf{v}^r_{n}=\left[e^{j(n-1)\mu_1^r} \quad e^{j(n-1)\mu_2^r}\quad \cdots\quad e^{j(n-1)\mu_{\hat{Z}}^r}\right]
\end{equation}
$\mathbf{v}^t_{m}$ is defined analogously. We combine the $NM$ equations in \eqref{CAE4} and solve for $\hat{\boldsymbol{\beta}}$ as
\begin{equation}
 \label{CAE7}
 \hat{\boldsymbol{\beta}}=(\hat{\mathbf{W}}_D^H\hat{\mathbf{W}}_D+\sigma\mathbf{I})\hat{\mathbf{W}}_D^H\hat{\mathbf{h}}
\end{equation}
where $\hat{\mathbf{h}}=\left[\hat{\mathbf{h}}_{11}^T\quad\hat{\mathbf{h}}_{12}\quad\cdots\quad\hat{\mathbf{h}}_{NM}\right]$ and $\hat{\mathbf{W}}_D=\begin{bmatrix}
                     \hat{\mathbf{W}}_{11}& 
                     \hat{\mathbf{W}}_{12}&
                     \cdots &
                     \hat{\mathbf{W}}_{NM}
                    \end{bmatrix}^T$. 
It should be noted that the choice of using \eqref{CAE3} or \eqref{CAE7} is essentially a compromise between complexity and accuracy, since the 
improved
amplitude estimates in \eqref{CAE7} is achieved at the cost of increased computational complexity. We will utilize \eqref{CAE7} for our analysis in this paper.
\subsubsection{Channel Prediction}
Once the parameters of the model have been estimated, the time-varying frequency selective channel is predicted via
\begin{equation}
 \label{CPD1}
 \tilde{\mathbf{H}}(q,k)=\sum_{z=1}^{\hat{Z}}\hat{\beta}_{z}\mathbf{a}_r(\hat{\mu}^{\mathrm{r}}_z)\mathbf{a}^T_t(\hat{\mu}^{\mathrm{t}}_z)\exp(jq\hat{\gamma}_{z}-jk\hat{\eta}_z)
\end{equation}
\subsection{TSSM-WIMEMCHAP}
Unlike the DOD/DOA-WIMEMCHAP approach, the TSSM-WIMEMCHAP involves 3D parameter estimation. The steps involved in the TSSM-WIMEMCHAP approach are
as follow.
\subsubsection{Covariance Matrix Estimation}
The covariance matrix, containing the receive spatial, temporal and frequency correlations, is estimated from \eqref{eq19} as
\begin{equation}
 \label{TSS1}
 \mathbf{C}_{\mathrm{t}}=\frac{1}{MRT}\hat{\mathbf{{X}}}_{\mathrm{t}}\hat{\mathbf{{X}}}_{\mathrm{t}}^H
\end{equation}
\subsubsection{Number of Paths Estimation and Subspace Decomposition}
We again estimate $Z$ using the MMSE-MDL criterion in \eqref{eq:eq30} and eigendecompose $ \mathbf{C}_{\mathrm{t}}$ as
\begin{equation}
 \label{TSS2}
\mathbf{C}_{\mathrm{t}} =\mathbf{E}_{\mathrm{s}}\Lambda_{\mathrm{s}}\mathbf{E}_{\mathrm{s}}^H+\mathbf{E}_{\mathrm{n}}\Lambda_{\mathrm{n}}\mathbf{E}_{\mathrm{n}}^H
\end{equation}
\subsubsection{Parameter Estimation}
Extraction of the parameter sets $\{\mu_z^r,\gamma_z,\eta_z\}_{z=1}^{\hat{Z}}$ requires a 3D estimation procedure. Similar to \eqref{eq:eqPPE3}, we form
the following 3D selection matrices
\begin{align}
 \label{TSS3}
 \mathbf{S}_{\theta 1}&=\mathbf{I}_{S}\otimes\mathbf{I}_{U}\otimes\mathbf{S}_{1\theta }\nonumber\\
 \mathbf{S}_{\theta 2}&=\mathbf{I}_{S}\otimes\mathbf{I}_{U}\otimes\mathbf{S}_{2\theta }\nonumber\\
 \mathbf{S}_{\gamma 1}&=\mathbf{I}_{S}\otimes\mathbf{S}_{1\gamma }\otimes\mathbf{I}_{N}\nonumber\\
 \mathbf{S}_{\gamma 2}&=\mathbf{I}_{S}\otimes\mathbf{S}_{2\gamma}\otimes\mathbf{I}_{N}\nonumber\\
 \mathbf{S}_{\eta 1}&=\mathbf{S}_{1\eta}\otimes\mathbf{I}_{U}\otimes\mathbf{I}_{N}\nonumber\\
 \mathbf{S}_{\eta 2}&=\mathbf{S}_{2\eta}\otimes\mathbf{I}_{U}\otimes\mathbf{I}_{N}
\end{align}
Using the selection matrices in \eqref{TSS3}, we form 3D invariance equations analogous to \eqref{eq:eqPPE4} and solve
\eqref{eq:eqPPE6b}--\eqref{eq:eqPPE6d}, \eqref{eq:eqPPE7}, \eqref{eq:eqPPE8}, \eqref{eq:eqPPE9b}--\eqref{eq:eqPPE9d} and \eqref{eq:eqPPE10b}--\eqref{eq:eqPPE10d} 
to obtain the parameter estimates.
\subsubsection{TSS Estimation}
The TSS can be similarly obtained via a least square approach. We assume that the TSS of the scattering sources are equal for all subcarriers and
use the channel for the first subcarrier in the estimation\footnote{The accuracy of the TSS estimation may be improved by incorporating all subcarriers. However,
the computational complexity will scale with the number of subcarriers.}.  Let $\mathbf{s}^m=[\mathbf{s}_1(m)\quad \mathbf{s}_2(m)\quad\cdots\quad\mathbf{s}_{\hat{Z}}(m)]^T$
be a vector containing the $m$th entry of the TSS for all paths. Using \eqref{eq9}, we obtain
\begin{equation}
 \label{TSSE1}
 \hat{\mathbf{h}}_m=\hat{\mathbf{W}}_m\mathbf{s}^m+\mathbf{n}\quad \forall m\in[1,M]
\end{equation}
where $\hat{\mathbf{h}}_m=[\hat{h}_{1m}(1),\hat{h}_{1m}(2),\cdots,\hat{h}_{1m}(Q),\cdots,\hat{h}_{Nm}(Q)]^T$ and $\hat{\mathbf{W}}_m$ is defined as
\begin{equation}
 \label{TSSE2}
 \hat{\mathbf{W}}_m=\mathbf{A}_{\mathrm{r}}\diamond\hat{\mathbf{W}}
\end{equation}
where $\mathbf{A}_{\mathrm{r}}$ is the receive array steering matrix defined for the ULA as
\begin{align}
 \label{TSSE3}
 \mathbf{A}_{\mathrm{r}}&=[\mathbf{a}(\mu^{\mathrm{r}}_1), \mathbf{a}(\mu^{\mathrm{r}}_2),\cdots,\mathbf{a}(\mu^{\mathrm{r}}_Z)]\nonumber\\
  &=\begin{bmatrix}
	      1 & 1 &\cdots & 1\\
	      e^{j\mu_1^r} & e^{j\mu_2^r} & \cdots & e^{j\mu_Z^r}\\
	      \vdots  &\vdots & \ddots & \vdots\\
	      e^{j(N-1)\mu_1^r} &e^{j(N-1)\mu_2^r} & \cdots & e^{j(N-1)\mu_Z^r}
    \end{bmatrix}
\end{align}
We solve \eqref{TSSE1} using the least square approach and estimate the TSS for  each path as
\begin{equation}
 \label{TSSE4}
 \hat{\mathbf{s}}_z=[\hat{\mathbf{s}}^1(z) \quad \hat{\mathbf{s}}^2(z) \quad \cdots \quad \hat{\mathbf{s}}^{M}(z)]^T
\end{equation}
\subsubsection{Channel Prediction}
Extrapolation of the channel using the TSSM-WIMEMCHAP is obtained from
\begin{equation}
 \label{TSSP1}
 \tilde{\mathbf{H}}(q,k)=\sum_{z=1}^{\hat{Z}}\mathbf{a}_r(\hat{\mu}^{\mathrm{r}}_z)\hat{\mathbf{s}}^T_z\exp(jq\hat{\gamma}_{z}-jk\hat{\eta}_z)
\end{equation}
\subsection{MSSM-WIMEMCHAP}
This approach involves estimation of the Doppler shifts and delays, which can be achieved via a 2D estimation procedure. This method is essentially
an extension of the SISO schemes in \cite{Liu2006} to MIMO channels. A summary of the steps in the
prediction is given below.
\subsubsection{Covariance Matrix Estimation}
The time-frequency covariance matrix is estimated using
\begin{equation}
 \label{MSSM1}
 \mathbf{C}_{\mathrm{m}}=\frac{1}{NMRT}\hat{\mathbf{X}}_{\mathrm{m}}\hat{\mathbf{X}}_{\mathrm{m}}^H
\end{equation}
\subsubsection{Number of Paths Estimation and Subspace Decomposition}
We estimate the number of paths using the MMSE-MDL criterion in \eqref{eq:eq30} with the eigenvalues of $ \mathbf{C}_{\mathrm{M}}$. The eigendecomposition of
$\mathbf{C}_{\mathrm{M}}$ can thus be expressed as
\begin{equation}
 \label{MSSM2}
\mathbf{C}_{\mathrm{M}} =\mathbf{E}_{\mathrm{s}}\Lambda_{\mathrm{s}}\mathbf{E}_{\mathrm{s}}^H+\mathbf{E}_{\mathrm{n}}\Lambda_{\mathrm{n}}\mathbf{E}_{\mathrm{n}}^H
\end{equation}
\subsubsection{Parameter Estimation}
Extraction of the parameter sets $\{\gamma_z,\eta_z\}_{z=1}^{\hat{Z}}$  requires 2D estimation and four selection matrices defined as
\begin{align}
 \label{MSSM3}
 \mathbf{S}_{\gamma 1}&=\mathbf{I}_{U}\otimes\mathbf{S}_{1\gamma }\nonumber\\
 \mathbf{S}_{\gamma 2}&=\mathbf{I}_{U}\otimes\mathbf{S}_{2\gamma}\nonumber\\
 \mathbf{S}_{\eta 1}&=\mathbf{S}_{1\eta}\otimes\mathbf{I}_{S}\nonumber\\
 \mathbf{S}_{\eta 2}&=\mathbf{S}_{2\eta}\otimes\mathbf{I}_{S}
\end{align}
We solve \eqref{eq:eqPPE6c}--\eqref{eq:eqPPE6d}, \eqref{eq:eqPPE7}--\eqref{eq:eqPPE8}, \eqref{eq:eqPPE9c}--\eqref{eq:eqPPE9d} and \eqref{eq:eqPPE10c}--\eqref{eq:eqPPE10d} 
to obtain the parameter estimates.
\subsubsection{MSS Estimation}
Let $\mathbf{s}_{nm}=[\mathbf{S}_1(n,m)\quad\mathbf{S}_2(n,m)\quad\cdots\quad\mathbf{S}_{Z}(n,m)]^T\in\mathbb{C}^{Z\times 1}$ be a vector containing  the
$(n,m)$th entry of the MSS for all paths. Using \eqref{eq10}, it can be easily shown that
\begin{equation}
 \label{MSSM4}
 \hat{\mathbf{h}}_{nm}=\hat{\mathbf{W}}_{11}\mathbf{s}_{nm}+\mathbf{n}
\end{equation}
for $n\in[1,N]$ and $m\in[1,M]$. We find the least square solution $\hat{\mathbf{s}}_{nm}$ to \eqref{MSSM4} for all antenna pairs and compute 
the MSS for the $z$th path as
\begin{equation}
 \label{MSSM5}
 \hat{\mathbf{S}}_z=\begin{bmatrix}
                     \hat{\mathbf{s}}_{11}(z) & \hat{\mathbf{s}}_{12}(z) & \cdots & \hat{\mathbf{s}}_{1M}(z)\\
                     \hat{\mathbf{s}}_{21}(z) & \hat{\mathbf{s}}_{22}(z) & \cdots & \hat{\mathbf{s}}_{2M}(z)\\
                     \vdots & \vdots & \ddots & \vdots\\
                     \hat{\mathbf{s}}_{N1}(z) & \hat{\mathbf{s}}_{N2}(z) & \cdots & \hat{\mathbf{s}}_{NM}(z)
                    \end{bmatrix}
\end{equation}
\subsubsection{Channel Prediction}
Channel prediction using the MSSM-WIMEMCHAP is achieved using
\begin{equation}
 \label{MSSM6}
 \tilde{\mathbf{H}}(q,k)=\sum_{z=1}^{\hat{Z}}\hat{\mathbf{S}}^T_z\exp(jq\hat{\gamma}_{z}-jk\hat{\eta}_z)
\end{equation}
\section{Performance Bounds}\label{sec:bound}
A commonly used bound on the performance of an unbiased estimator is the Cramer-Rao lower bound (CRLB) \cite{Kay1}. In this section, 
we derive the bound on the variance of prediction error in wideband MIMO systems. While similar results have been presented in 
\cite{Larsen2008,Larsen2009}, we present an alternative, simpler formulation.
Consider the vectorized model in \eqref{eq11}. Using the properties of Kronecker products, $\mathbf{h}(q,k)$ can be written
as
\begin{equation}
 \label{PEB1}
 \mathbf{h}(q,k)=\left(\mathbf{A}_{\mathrm{r}}\diamond\mathbf{A}_{\mathrm{t}}\right)\boldsymbol{\alpha}(q,k)
\end{equation}
where
\begin{equation}
 \label{PEB2}
 \boldsymbol{\alpha}(q,k)=\left[\beta_1\exp(jq\gamma_1-jk\eta_1)\,\cdots\,\beta_Z\exp(jq\gamma_Z-jk\eta_Z)\right]
\end{equation}
and $\mathbf{A}_{\mathrm{r}}$ is the receive array steering matrix in \eqref{TSSE3}. The transmit array response matrix,
$\mathbf{A}_{\mathrm{t}}$ is defined analogously by replacing $\mu^{\mathrm{r}}_z$  with $\mu^{\mathrm{t}}_z$. We arrange the $QK$ known samples 
into the vector
\begin{align}
 \label{PEB4}
 \mathbf{h}&=\left[\mathbf{h}^T(1,1),\cdots,\mathbf{h}^T(Q,1),\mathbf{h}^T(1,2),\cdots,\right.\nonumber\\
 &\qquad\left.\mathbf{h}^T(Q,2),\cdots,\mathbf{h}^T(1,3),\cdots,\mathbf{h}^T(Q,K)\right]
\end{align}
and define Vandermonde matrices $\mathbf{A}_{\mathrm{d}}$ and $\mathbf{A}_{\mathrm{f}}$ as
\begin{equation}
 \label{PEB5}
 \mathbf{A}_{\mathrm{d}}=\begin{bmatrix}
             1  & \cdots & 1\\
             \exp(j\gamma_1)  & \cdots & \exp(j\gamma_Z)\\
             \vdots  & \ddots & \vdots\\
             \exp(j(Q-1)\gamma_1) & \cdots & \exp(j(Q-1)\gamma_Z)
            \end{bmatrix}
\end{equation}
and
\begin{equation}
 \label{PEB6}
 \mathbf{A}_{\mathrm{f}}=\begin{bmatrix}
             1  & \cdots & 1\\
             \exp(-j\eta_1)  & \cdots & \exp(-j\eta_Z)\\
             \vdots  & \ddots & \vdots\\
             \exp(-j(K-1)\eta_1) & \cdots & \exp(-j(K-1)\eta_Z)
            \end{bmatrix}
\end{equation}
It can easily be verified that
\begin{align}
 \label{PEB7}
 \mathbf{h}&=\left(\mathbf{A}_{\mathrm{r}}\diamond\mathbf{A}_{\mathrm{t}}\diamond\mathbf{A}_{\mathrm{d}}\diamond\mathbf{A}_{\mathrm{f}}\right)\boldsymbol{\beta}\nonumber\\
 &=\mathbf{A}\boldsymbol{\beta}
\end{align}
where $\mathbf{A}=(\mathbf{A}_{\mathrm{r}}\diamond\mathbf{A}_{\mathrm{t}}\diamond\mathbf{A}_{\mathrm{d}}\diamond\mathbf{A}_{\mathrm{f}})$. Let the
parameterization of the channel be denoted by
\begin{equation}
 \boldsymbol{\Theta}=\left[\sigma^2,\boldsymbol{\theta}^T,\boldsymbol{\phi}^T,\boldsymbol{\gamma}^T,\boldsymbol{\eta}^T,\mathfrak{R}(\boldsymbol{\beta}^T)
 ,\mathfrak{I}(\boldsymbol{\beta}^T)\right]^T
\end{equation}
It should be noted that our channel models represent continuous non--linear functions of the parameters and the bound on estimation/prediction of 
$\mathbf{h}(q,k)$
can be found using the vector formulation of the Cramer--Rao bound for function of parameters. The prediction error can therefore be 
bounded by \cite{Kay1}
\begin{align}
 \label{eq:eq14}
 \mathbf{C}_e(q,k) &=\mathbb{E}[(\mathbf{\hat{h}}(q,k)-\mathbf{h}(q,k))(\mathbf{\hat{h}}(q,k)-\mathbf{h}(q,k))^H]\nonumber\\
 &\geq\frac{\partial \mathbf{h}(q,k)}{\partial \boldsymbol{\Theta}}^H\mathfrak{{F}}^{-1}(\boldsymbol{\Theta})\frac{\partial \mathbf{h}(q,k)}{\partial \boldsymbol{\Theta} }
\end{align}
where $\mathfrak{F}^{-1}(\boldsymbol{\Theta})$ is the lower bound on the estimation of channel parameters, and $\mathfrak{F}(\boldsymbol{\Theta})$ 
is the Fisher information matrix (FIM), calculated using Bangs formula \cite{Kay1}
\begin{equation}
\label{eq:ex1}
\left[\mathfrak{F}(\boldsymbol{\Theta})\right]_{ij}=\operatorname{Tr}\left[\mathbf{C}^{-1}\frac{\partial \mathbf{C}}{\partial \boldsymbol{\Theta}_i}\mathbf{C}^{-1}\frac{\partial \mathbf{C}}{\partial \boldsymbol{\Theta}_j}\right]+2\mathfrak{R}\left[\frac{\partial \mathbf{h}^H}{\partial \boldsymbol{\Theta}_i}\mathbf{C}^{-1}\frac{\partial \mathbf{h}}{\partial \boldsymbol{\Theta}_j}\right]
\end{equation}
After evaluating the derivatives and performing straightforward but tedious simplications, $\mathfrak{F}(\boldsymbol{\Theta})$ is obtained as
\begin{equation}
 \mathfrak{F}(\boldsymbol{\Theta})=\begin{bmatrix}
                                  \frac{KQNM}{\sigma^4} & \mathbf{0}\\
                                  \mathbf{0}^T & \mathbf{J}(\boldsymbol{\Theta})
                                 \end{bmatrix}
\end{equation}
with the Fisher information submatrix defined as
\begin{align}
 \label{eq:eq40}
 \left[\mathbf{J}(\Theta)\right] &=\frac{2}{\sigma^2}\mathfrak{R}\left[(\mathbf{G}_5^H\mathbf{G}_5)\odot(\mathbf{G}_4^H\mathbf{G}_4)\odot(\mathbf{G}_3^H\mathbf{G}_3)
 \right.\nonumber\\ &\quad\quad\quad\left. \odot(\mathbf{G}_2^H\mathbf{G}_2)
 \odot(\mathbf{G}_1^H\mathbf{G}_1)\right]
\end{align}
where $\odot$ denotes the Hardamad product and $\mathbf{G}_1$--$\mathbf{G}_5$ are defined as
\begin{align}
 \label{eqP11}\mathbf{G}_1=&\left[\boldsymbol{\alpha}^T\quad\boldsymbol{\alpha}^T\quad\boldsymbol{\alpha}^T\quad\boldsymbol{\alpha}^T\quad\mathbf{1}^1\quad j\mathbf{1}^T\right]\\
 \label{eqP12}\mathbf{G}_2=&\left[\mathbf{D}_{\mathrm{r}}\quad \mathbf{A}_{\mathrm{r}}\quad \mathbf{A}_{\mathrm{r}}\quad \mathbf{A}_{\mathrm{r}}\quad \mathbf{A}_{\mathrm{r}}\quad \mathbf{A}_{\mathrm{r}}\right]\\
 \label{eqP13}\mathbf{G}_3=&\left[\mathbf{A}_{\mathrm{t}}\quad \mathbf{D}_{\mathrm{t}}\quad \mathbf{A}_{\mathrm{t}}\quad \mathbf{A}_{\mathrm{t}}\quad \mathbf{A}_{\mathrm{t}}\quad \mathbf{A}_{\mathrm{t}}\right]\\
 \label{eqP14}\mathbf{G}_4=&\left[\mathbf{A}_{\mathrm{d}}\quad \mathbf{A}_{\mathrm{d}}\quad \mathbf{D}_{\mathrm{d}}\quad \mathbf{A}_{\mathrm{d}}\quad \mathbf{A}_{\mathrm{d}}\quad \mathbf{A}_{\mathrm{d}}\right]\\
 \label{eqP15}\mathbf{G}_5=&\left[\mathbf{A}_{\mathrm{f}}\quad \mathbf{A}_{\mathrm{f}}\quad \mathbf{A}_{\mathrm{f}}\quad \mathbf{D}_{\mathrm{f}}\quad \mathbf{A}_{\mathrm{f}}\quad \mathbf{A}_{\mathrm{f}}\right]
\end{align}
The matrix $\mathbf{D}_{\mathrm{r}}\in\mathbb{C}^{N\times Z}$ is given by
\begin{align}
 \label{eq:eq19}
 \mathbf{D}_{\mathrm{r}}&=\left[\frac{\partial \mathbf{a}_{\mathrm{r}}(\mu^{\mathrm{r}}_1)}{\partial \mu^{\mathrm{r}}_1}\quad\cdots\frac{\partial \mathbf{a}_{\mathrm{r}}(\mu^{\mathrm{r}}_Z)}{\partial \mu^{\mathrm{r}}_Z}\right]\nonumber\\
 &=-j\mathbf{Z}_N\mathbf{A}_{\mathrm{r}}
\end{align}
where $\mathbf{Z}_k$ is a diagonal matrix
\begin{equation}
 \label{eq:eq20}
 \mathbf{Z}_k=\begin{bmatrix}
               0 & 0 & 0 & \cdots & 0\\
               0 & 1 & 0 & \cdots & 0\\
               0 & 0 & 2 & \cdots & 0\\
               \vdots&\vdots&\vdots&\ddots&\vdots\\
               0 & 0 & 0 & \cdots & k-1
              \end{bmatrix}
\end{equation}
$ \mathbf{D}_{\mathrm{t}}$, $ \mathbf{D}_{\mathrm{d}}$ and $ \mathbf{D}_{\mathrm{f}}$ are defined analogously to \eqref{eq:eq19}. Details of the derivations can be found in the Appendix.

\section{Numerical Simulations}\label{sec:simu}
In this section, we evaluate the performance of the proposed schemes. After describing the performance metrics and simulation parameters,
we evaluate the the parameter estimation accuracy, followed by the overall prediction performance. 
\subsection{Performance Metrics}
The overall performance of the prediction schemes is evaluated using Monte Carlo simulations with synthetic data and compared with
prediction error bounds in Section~\ref{sec:bound}. The normalized mean square error (NMSE) is used as the performance metric. 
We first define the normalized square error (NSE) over a single realization of the channel as
\begin{equation}
 \label{PerfM1}
 \mathrm{NSE}(q,k)=\frac{||\tilde{\mathbf{H}}(q,k)-\mathbf{H}(q,k)||^2_F}{\mathbb{E}\left[||\mathbf{H}(q,k)||^2_F\right]}
\end{equation}
where $||\cdot||_F$ denotes the Frobenious norm of the associated matrix. The expectation in \eqref{PerfM1} is approximated over the 
available temporal and frequency samples. 

The performance of the parameter estimation stage is evaluated in terms of the root mean square error (RMSE) defined, for a generic variable $x$, as
\begin{equation}
\label{PerfM2}
\mathrm{RMSE}(\hat{x})=\sqrt{\frac{1}{N_{\mathrm{c}}}\sum_{c=1}^{N_{\mathrm{c}}}(x-\hat{x}_c)^2}
\end{equation}
where $N_{\mathrm{c}}$ denotes the number of channel and/or noise realizations and $\hat{x}_c$ is the estimate of $x$ during the $c$th 
realization. The NMSE and RMSE are obtained by averaging \eqref{PerfM1} and \eqref{PerfM2} over $500$ realizations of the channel and/or noise. The 
RMSE is compared with the square root of the CRB (i.e., diagonal entries of the inverse of \eqref{eq:eq40}). 
\subsection{Simulation Parameters}
We consider a MIMO--OFDM system with bandwidth $B=20\,$MHz, $N_{\mathrm{sc}}=1024$  subcarriers including $K=64$ equally spaced pilot subcarriers. The 
transmit and receive antenna arrays are both 2-element arrays with inter-antenna spacing $d_{\mathrm{r}}=d_{\mathrm{t}}=\lambda/2$. 
We consider a carrier frequency of $f_{\mathrm{c}}=2.1\,$GHz and mobile velocity of $v=50\,$km/h. Except where otherwise stated, we use $50$ samples 
at a sampling rate of $10/\lambda$. We consider two methods of generating the channel parameters $\beta_z,\theta_z,\phi_z,\nu_z$. In simulation 
Scenario I, these are fixed to values given in Table~\ref{tab:tab3} for all realizations of $\mathbf{H}$ in \eqref{eq7}. In simulation Scenario II, 
they are randomly generated for each channel realization. The amplitudes  are generated as complex 
Gaussian distributed random variables, $\beta_z\sim\mathcal{CN}(0,1)$. The angles of arrival and departure are assumed to be uniformly distributed, 
i.e., $\theta_z,\phi_z\sim\mathcal{U}[-\pi,\pi)$. In both cases, the path delays are selected from the Urban 
macro (UMA) scenario in the WINNER II/3GPP channel \cite{WINNER}, given in Table~\ref{tab:tab3}. Unless otherwise stated, we consider a 6-path channel
with parameters in Table~\ref{tab:tab3} and the error is averaged over 500 noise realizations. . 
\subsection{Parameter Estimation Performance}

\begin{figure}
\centering
\includegraphics[height= 0.6\columnwidth,width=0.8\columnwidth]{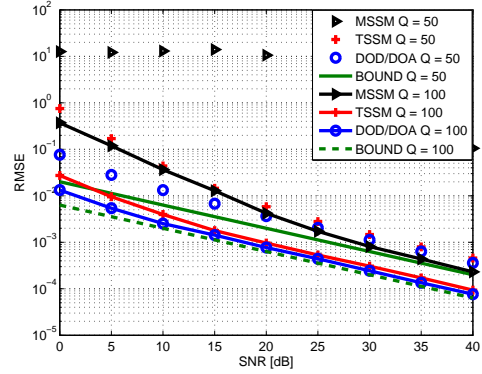}
\caption{The RMSE of Doppler frequency estimates versus SNR with $[50,100]$ known channel samples.}
\label{fig:fig2}
\end{figure}

\begin{figure}
\centering
\includegraphics[height=0.6\columnwidth,width=0.8\linewidth]{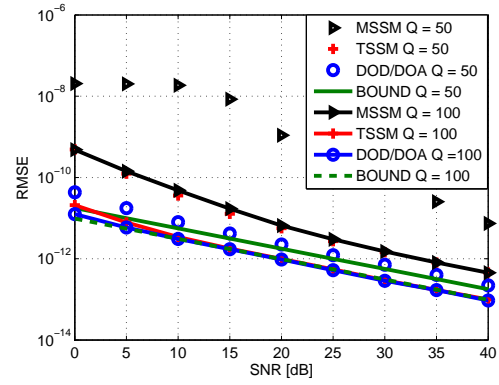}
\caption{The RMSE of delay estimates versus SNR with $[50,100]$ known channel samples.}
\label{fig:fig2a}
\end{figure}

Since Doppler frequency and delay estimation are part of all the methods, we present results showing the accuracy of their 
estimates in each algorithm.  Figure~\ref{fig:fig2} presents the RMSE of Doppler estimates versus 
SNR with 50 and 100 known samples of the channel. We observe that the performances of the three methods improves with increasing number 
of samples and approaches the bound as the SNR increases. Also, we note that the DOD/DOA method outperforms the TSSM and MSSM methods 
at all SNR values and that the MSSM method yields the highest RMSE. A possible reason for this is the additional channel structure 
revealed by sampling in a higher number of dimensions. Similar observations are made in Fig.~\ref{fig:fig2a}, where we plot the RMSE of delay 
estimates versus SNR.

\begin{table}[!t]
\renewcommand{\arraystretch}{1.5}
\caption{Propagation Channel Parameters (Scenario I)}
\label{tab:tab3}
\centering
\begin{tabular}{|c||c|c|c|c|c|}
\hline
Path  & \multicolumn{5}{|c|}{Parameters}\\
\hline
 & $\beta_z$ & $\theta_z$ & $\phi_z$ & $\tau_z$ (ns) & $\nu_z$(rad/s)\\
 \hline
1 & -0.76+0.074j & 0.49 & -2.90 & 0 & 185.10 \\
\hline
2 & -0.76+0.30j & -1.89 & 0.99 & 60 & -462.10\\
\hline
3 & -1.41+0.14j & -2.48 & 2.99 & 75 & 497.31\\
\hline
4 & 0.16-1.15j & -1.88 & 1.46 & 145 & -331.90\\
\hline
5 & 0.37-0.82j & -2.66 & 2.05 & 150 & 208.61\\
\hline
6 & -0.33+1.04j & -0.02 & -1.60 & 155 & -156.92\\
\hline
\end{tabular}
\end{table}
\subsection{Prediction Performance}
\begin{figure}
\centering
\includegraphics[height=0.6\columnwidth,width=0.8\linewidth]{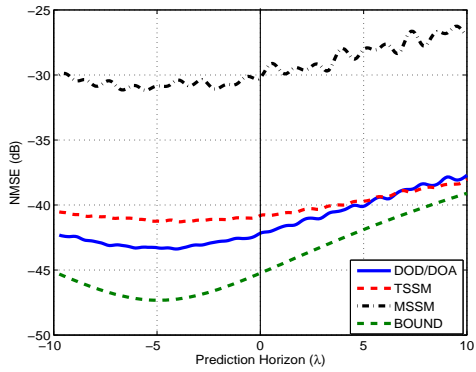}
\caption{The averaged NMSE versus prediction horizon for a $2\times 2$ MIMO channel prediction at $\mathrm{SNR}=15\,$dB.}
\label{fig:fig3}
\end{figure}

\begin{figure}
\centering
\includegraphics[height=0.6\columnwidth,width=0.8\linewidth]{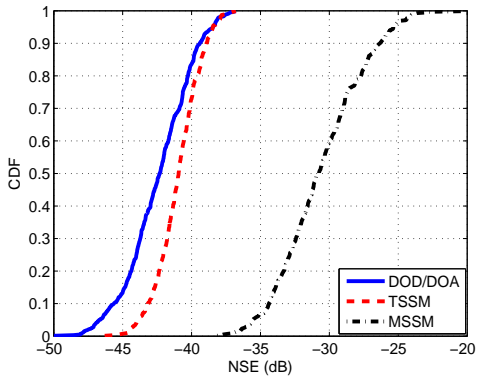}
\caption{The cummulative distribution function of NSE for a prediction horizon of $1\,\lambda$ at $\mathrm{SNR}=15\,$dB.}
\label{fig:fig3a}
\end{figure}

\begin{figure}
\centering
\includegraphics[height=0.6\columnwidth,width=0.8\linewidth]{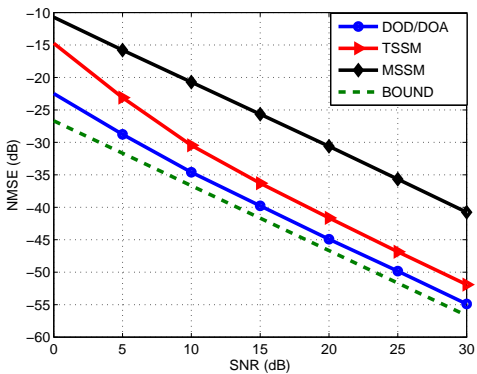}
\caption{The averaged NMSE versus SNR for a prediction horizon of $1\,\lambda$.}
\label{fig:fig4}
\end{figure}

We now evaluate the overall prediction performance of the proposed methods. Figures~\ref{fig:fig3}--\ref{fig:fig4} correspond to simulation Scenario I.
Figure~\ref{fig:fig3} presents the NMSE versus prediction horizon (in wavelengths) at an SNR of 15\,dB. The negative values of the prediction horizon
correspond to the estimation stage. Again, we observe that the DOD/DOA outperforms TSSM and MSSM methods. 
In Fig.~\ref{fig:fig3a}, we plot the corresponding cumulative distribution function (CDF) of the normalized square error (NSE) at
a prediction interval of $1\lambda$. The DOD/DOA method has the lowest NSE for all realizations followed by the TSSM. We also observe that 
utilizing the spatial information in parameter estimation and prediction in the DOD/DOA and TSSM results in a decrease of about 12\,dB relative 
to MSSM.

The effects of increasing SNR on the performance of the proposed schemes is shown in Fig.~\ref{fig:fig4} where we plot the NMSE versus SNR for 
a prediction horizon of $1\lambda$. As expected, as a consequence of improved parameter estimation, the performance of the algorithms improves with 
increasing SNR. We observe that the performance of the DOD/DOA and TSSM methods approaches the bound as SNR increases with 
the DOD/DOA having the lowest NMSE over the entire SNR range considered. This agrees with observations in \cite{Larsen2009} where it was shown 
that the prediction error bound obtained from the DOD/DOA model is lower than that for the vector spatial signature model.

\begin{figure}
\centering
\includegraphics[height=0.6\columnwidth,width=0.8\linewidth]{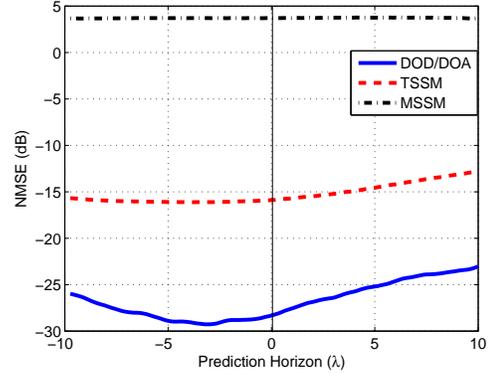}
\caption{The averaged NMSE versus prediction horizon for a $2\times 2$ MIMO channel prediction at $\mathrm{SNR}=15\,$dB. 
Channel and noise are random for each realization. The bound falls outside the plotted range.}
\label{fig:fig5}
\end{figure}

\begin{figure}
\centering
\includegraphics[height=0.6\columnwidth,width=0.8\linewidth]{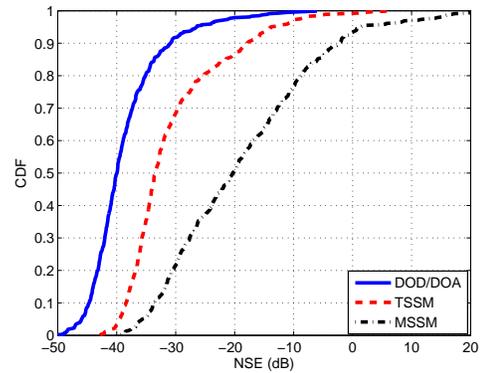}
\caption{The cummulative distribution function of NSE for a prediction horizon of $1\,\lambda$ at $\mathrm{SNR}=15\,$dB. Channel and noise are random for each realization.}
\label{fig:fig5a}
\end{figure}

We now present results for simulation Scenario II.  In Fig.~\ref{fig:fig5}, we present the NMSE versus prediction horizon at an SNR of 15\,dB. Here,
we 
observe that the averaged performance of all methods degrades when compared to Scenario I. However, the performance of the DOD/DOA and TSSM schemes
are still reasonable with a maximum NMSE of about -12\,dB for the TSSM and -22\,dB for the DOD/DOA over the $15\lambda$ prediction horizon shown. 
We observe that the MSSM performs poorly with an NMSE of approximately 4.8\,dB over the entire region considered. A possible explanation for the
increase in NMSE is that unlike in simulation Scenario I, certain channel realizations have parameters which are closer than the resolution limit of 
the parameter estimation stage, leading to reduced parameter estimation and prediction performance. 
The CDF of the NSE corresponding to the mean results in Fig.~\ref{fig:fig5} at a prediction interval of $1\lambda$ is shown in Fig.~\ref{fig:fig5a}. 

\begin{figure}
\centering
\includegraphics[height=0.6\columnwidth,width=0.8\linewidth]{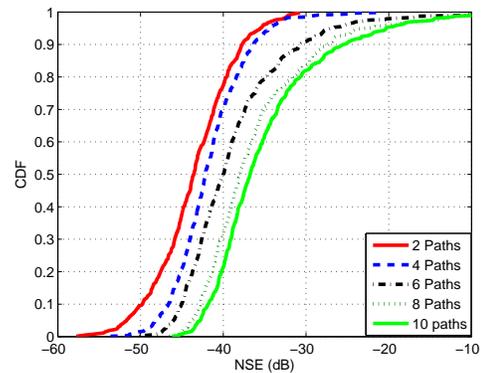}
\caption{The cummulative distribution function of NSE for a prediction horizon of $1\,\lambda$ at $\mathrm{SNR}=15\,$dB with $Z=[2,4,6,8,10]\,$ paths.}
\label{fig:fig6}
\end{figure}

Finally, the CDF of the prediction error of the DOD/DOA for different number of propagation paths is presented in Fig.~\ref{fig:fig6}. We observe that the performance of the algorithm degrades with increasing number of paths. 


\section{Conclusion}\label{sec:conc}
We have presented three different parametric schemes for the prediction of mobile  MIMO-OFDM channels. 
The predictors are based on different formulations of the double directional model and original adaptation of multidimensional ESPRIT to
jointly extract the channel parameters. Using the vector formulation of Cramer Rao bound for functions of parameters, a simplified 
form of the lower bound on prediction error in MIMO-OFDM channel was derived.  Numerical simulations indicate that the   
performance of the algorithms approaches the error bound with increasing SNR and/or number of samples. We have quantified the parameter estimation
and channel prediction improvement afforded by the spatial structure of the channel revealed by multiple sampling of the wavefield. 
The method utilizing both transmit and receive spatial  information (DOD/DOA-WIMEMCHAP) outperform those with only receive spatial information (TSSM-WIMEMCHAP) 
and no spatial information (MSSM-WIMEMCHAP).

\vspace{10pt}
\appendices
\section{Derivatives for the FIM}\label{sec:ApA}
In this section, we give the derivatives of the covariance matrix and observation vector $\mathbf{h}$ required for evaluating the FIM expression 
as given in \eqref{eq:eq40}.
\subsection{Derivative with Respect $\sigma^2$}
Based on the independent Gaussian noise assumption, the derivative of the covariance matrix $\mathbf{C}=\sigma^2\mathbf{I}$ is
\begin{equation}\label{eqC1}
\frac{\partial \mathbf{C}}{\partial \sigma^2}=\mathbf{I}
\end{equation}
On substituting \eqref{eqC1} into \eqref{eq:ex1}, the entry of the FIM dependent on noise variance is obtained as
\begin{equation}
\operatorname{Tr}\left[\mathbf{C}^{-1}\frac{\partial \mathbf{C}}{\partial \boldsymbol{\Theta}}\mathbf{C}^{-1}\frac{\partial \mathbf{C}}{\partial \boldsymbol{\Theta}}\right]=\frac{KQNM}{\sigma^4}
\end{equation}
\subsection{Derivative With Respect to $\mathfrak{R}(\boldsymbol{\beta})$}
\begin{align}
\label{eq:eqAP1}
\frac{\partial \mathbf{h}}{\partial \mathfrak{R}({\beta_z})}&=\frac{\partial (\mathbf{A}_{\mathrm{r}}\diamond\mathbf{A}_{\mathrm{t}}
\diamond\mathbf{A}_{\mathrm{d}}\diamond\mathbf{A}_{\mathrm{f}})\boldsymbol{\beta}}{\partial \mathfrak{R}({\beta_z})}\nonumber\\
&=(\mathbf{A}_{\mathrm{r}}\diamond\mathbf{A}_{\mathrm{t}}\diamond\mathbf{A}_{\mathrm{d}}\diamond\mathbf{A}_{\mathrm{f}})
\frac{\partial \boldsymbol{\beta}}{\partial \beta_z}\nonumber\\
&=(\mathbf{A}_{\mathrm{r}}\diamond\mathbf{A}_{\mathrm{t}}\diamond\mathbf{A}_{\mathrm{d}}\diamond\mathbf{A}_{\mathrm{f}})\boldsymbol{\Pi}_z\nonumber\\
\end{align}
where $\boldsymbol{\Pi}_z$ is a $Z\times 1$ vector having a 1 as the $z$th element and all other elements zero. \eqref{eq:eqAP1} can further be simplified to
\begin{align}
\label{eq:eqAP2}
\frac{\partial \mathbf{h}}{\partial \mathfrak{R}({\beta_z})}&=[\mathbf{A}_{\mathrm{r}}]_{:,z}\otimes[\mathbf{A}_{\mathrm{t}}]_{:,z}\otimes
[\mathbf{A}_{\mathrm{d}}]_{:,z}\otimes[\mathbf{A}_{\mathrm{f}}]_{:,z}\nonumber\\
&=[\mathbf{A}_{\mathrm{r}}\diamond\mathbf{A}_{\mathrm{t}}\diamond\mathbf{A}_{\mathrm{d}}\diamond\mathbf{A}_{\mathrm{f}}]_{:,z}\nonumber\\
&=[\mathbf{A}]_{:,z}
\end{align}
where $[\mathbf{B}]_{:,z}$ denotes the $z$th column of $\mathbf{B}$. Using \eqref{eq:eqAP2}, the derivative with respect to  $\mathfrak{R}(\boldsymbol{\beta})$ is
\begin{align}
\label{eq:eqAP3}
\frac{\partial \mathbf{h}}{\partial \mathfrak{R}(\boldsymbol{\beta})}&=\left[\frac{\partial \mathbf{h}}{\partial \mathfrak{R}({\beta_1})}\quad \frac{\partial \mathbf{h}}{\partial \mathfrak{R}({\beta_2})}\quad\cdots\quad\frac{\partial \mathbf{h}}{\partial \mathfrak{R}({\beta_Z})}\right]\nonumber\\
&=\left[[\mathbf{A}]_{:,1}\quad\cdots\quad[\mathbf{A}]_{:,Z}\right]\nonumber\\
&=\mathbf{A}
\end{align}
\subsection{Derivative With Respect to $\mathfrak{I}(\boldsymbol{\beta})$}
\begin{align}
\label{eq:eqAP4}
\frac{\partial \mathbf{h}}{\partial \mathfrak{I}({\beta_k})}&=(\mathbf{A}_{\mathrm{r}}\diamond\mathbf{A}_{\mathrm{t}}\diamond\mathbf{A}_{\mathrm{d}}
\diamond\mathbf{A}_{\mathrm{f}})\frac{\partial \boldsymbol{\beta}}{\partial \mathfrak{I}(\beta_z)}\nonumber\\
&=j(\mathbf{A}_{\mathrm{r}}\diamond\mathbf{A}_{\mathrm{t}}\diamond\mathbf{A}_{\mathrm{d}}\diamond\mathbf{A}_{\mathrm{f}})\boldsymbol{\Pi}_Z\nonumber\\
&=j\mathbf{A}
\end{align}
\subsection{Derivative With Respect to $\boldsymbol{\mu}^{\mathrm{r}}$}
\begin{align}
\label{eq:eqAP5}
\frac{\partial \mathbf{h}}{\partial \mu^{\mathrm{r}}_z}&=\left(\frac{\partial \mathbf{A}_{\mathrm{r}}}{\partial \mu^{\mathrm{r}}_z} \diamond\mathbf{A}_{\mathrm{t}}\diamond\mathbf{A}_{\mathrm{d}}
\diamond\mathbf{A}_{\mathrm{f}}\right)\boldsymbol{\beta}\nonumber\\
&=\left(\frac{\partial[\mathbf{A}_{\mathrm{r}}]_{:,z}}{\partial \mu^{\mathrm{r}}_z}\otimes[\mathbf{A}_{\mathrm{t}}]_{:,z}\otimes[\mathbf{A}_{\mathrm{d}}]_{:,z}\otimes[\mathbf{A}_{\mathrm{f}}]_{:,z}\right)\beta_z
\end{align}
Using \eqref{eq:eqAP5}, we obtain
\begin{align}
\label{eq:eqAP6}
\frac{\partial \mathbf{h}}{\partial \boldsymbol{\mu}^{\mathrm{r}}}&=\left[\frac{\partial \mathbf{h}}{\partial \mu^{\mathrm{r}}_1}\quad\frac{\partial \mathbf{h}}{\partial \mu^{\mathrm{r}}_2}\quad\cdots\quad\frac{\partial \mathbf{h}}{\partial \mu^{\mathrm{r}}_Z}\right]\nonumber\\
&=\left[\left(\frac{\partial[\mathbf{A}_{\mathrm{r}}]_{:,1}}{\partial\mu^{\mathrm{r}}_1}\otimes[\mathbf{A}_{\mathrm{t}}]_{:,1}\otimes[\mathbf{A}_{\mathrm{d}}]_{:,1}\otimes[\mathbf{A}_{\mathrm{f}}]_{:,1}\right)\beta_1\right.\quad\cdots\nonumber\\
&\qquad\left.\left(\frac{\partial[\mathbf{A}_{\mathrm{r}}]_{:,Z}}{\partial \mu^{\mathrm{r}}_Z}\otimes[\mathbf{A}_{\mathrm{t}}]_{:,Z}\otimes[\mathbf{A}_{\mathrm{d}}]_{:,Z}\otimes[\mathbf{A}_{\mathrm{f}}]_{:,Z}\right)\beta_Z\right]\nonumber\\
&=(\mathbf{D}_{\mathrm{r}}\diamond\mathbf{A}_{\mathrm{t}}\diamond\mathbf{A}_{\mathrm{d}}\diamond\mathbf{A}_{\mathrm{f}})\mathbf{X}
\end{align}
where
\begin{equation}
\mathbf{D}_r=\left[\frac{\partial[\mathbf{A}_r]_{:,1}}{\partial\mu^{\mathrm{r}}_1}\quad\frac{\partial[\mathbf{A}_r]_{:,2}}{\partial\mu^{\mathrm{r}}_2}\quad\cdots\quad\frac{\partial[\mathbf{A}_r]_{:,Z}}{\partial\mu^{\mathrm{r}}_Z}\right]
\end{equation}
and $\mathbf{X}$ is a diagonal matrix with the complex amplitudes $\boldsymbol{\beta}$ on its diagonal. For the ULA, $\mathbf{D}_r$ can be found using \eqref{TSSE3} to be
\begin{equation}\label{eq:eqAP7}
\mathbf{D}_{\mathrm{r}}=\begin{bmatrix}
0 & \cdots & 0\\
je^{j\mu^{\mathrm{r}}_1}  & \cdots & je^{j\mu^{\mathrm{r}}_Z}\\
2je^{j2\mu^{\mathrm{r}}_1}  & \cdots & 2je^{j2\mu^{\mathrm{r}}_Z}\\
\vdots & \ddots & \vdots\\
j(N-1)e^{j(N-1)\mu^{\mathrm{r}}_1}  & \cdots & j(N-1)e^{j(N-1)\mu^{\mathrm{r}}_Z}
\end{bmatrix}
\end{equation}
which can be expressed in terms of $\mathbf{A}_{\mathrm{r}}$ as
\begin{equation}
\mathbf{D}_{\mathrm{r}}=j\begin{bmatrix}
0 & 0 & \cdots & 0\\
0 & 1 & \cdots & 0\\
\vdots & \vdots & \ddots & \vdots\\
0 & 0 & \cdots & (N-1)
\end{bmatrix}\mathbf{A}_{\mathrm{r}}
\end{equation}

\subsection{Derivative With Respect to $\boldsymbol{\mu}^{\mathrm{t}}$}
\begin{align}
\label{eq:eqAP8}
\frac{\partial \mathbf{h}}{\partial \boldsymbol{\mu}^{\mathrm{t}}}&=\left[\frac{\partial \mathbf{h}}{\partial \mu^{\mathrm{t}}_1}\quad\frac{\partial \mathbf{h}}{\partial \mu^{\mathrm{t}}_2}\quad\cdots\quad\frac{\partial \mathbf{h}}{\partial \mu^{\mathrm{t}}_Z}\right]\nonumber\\
&=\left[\left([\mathbf{A}_r]_{:,1}\otimes\frac{\partial[\mathbf{A}_t]_{:,1}}{\partial\mu^{\mathrm{t}}_1}\otimes[\mathbf{A}_d]_{:,1}\otimes[\mathbf{A}_f]_{:,1}\right)\beta_1\right.\quad\cdots\nonumber\\
&\qquad\left.\left([\mathbf{A}_r]_{:,Z}\otimes\frac{\partial[\mathbf{A}_t]_{:,Z}}{\partial \mu^{\mathrm{t}}_Z}\otimes[\mathbf{A}_d]_{:,Z}\otimes[\mathbf{A}_f]_{:,Z}\right)\beta_Z\right]\nonumber\\
&=(\mathbf{A}_r\diamond\mathbf{D}_t\diamond\mathbf{A}_d\diamond\mathbf{A}_f)\mathbf{X}
\end{align}
where $\mathbf{D}_{\mathrm{t}}$ is defined analogous to $\mathbf{D}_{\mathrm{r}}$.
\subsection{Derivative With Respect to $\boldsymbol{\nu}$}
\begin{align}
\label{eq:eqAP9}
\frac{\partial \mathbf{h}}{\partial \boldsymbol{\nu}}&=\left[\frac{\partial \mathbf{h}}{\partial \nu_1}\quad\frac{\partial \mathbf{h}}{\partial \nu_2}\quad\cdots\quad\frac{\partial \mathbf{h}}{\partial \nu_Z}\right]\nonumber\\
&=\left[\left([\mathbf{A}_r]_{:,1}\otimes[\mathbf{A}_t]_{:,1}\otimes\frac{\partial[\mathbf{A}_d]_{:,1}}{\partial\nu_1}\otimes[\mathbf{A}_f]_{:,1}\right)\beta_1\right.\quad\cdots\nonumber\\
&\qquad\left.\left([\mathbf{A}_r]_{:,Z}\otimes[\mathbf{A}_t]_{:,Z}\right)\beta_Z\otimes\frac{\partial[\mathbf{A}_d]_{:,Z}}{\partial \nu_Z}\otimes[\mathbf{A}_f]_{:,Z}\right]\nonumber\\
&=(\mathbf{A}_r\diamond\mathbf{A}_t\diamond\mathbf{D}_d\diamond\mathbf{A}_f)\mathbf{X}
\end{align}
where $\mathbf{D}_d$ is defined similar to \eqref{eq:eqAP7}.
\subsection{Derivative With Respect to $\boldsymbol{\eta}$}

Following a procedure similar to those presented above, the derivative of $\mathbf{h}$ with respect to $\boldsymbol{\eta}$ can be shown to be
\begin{equation}
\frac{\partial \mathbf{h}}{\partial \boldsymbol{\eta}}=(\mathbf{A}_r\diamond\mathbf{A}_t\diamond\mathbf{A}_d\diamond\mathbf{D}_f)\mathbf{X}
\end{equation}
\section{Evaluation of FIM and Error Bound}
Once the derivatives of the channel observation $\mathbf{h}$ and the covariance matrix have been obtained, the expression for the FIM and the prediction error bound are obtained by substituting the derivatives in Appendix~\ref{sec:ApA}. Using \eqref{PEB4} and \eqref{eq:ex1}, the matrix $\mathbf{J}$ in \eqref{eq:eq40} is obtained as
\begin{equation}
\label{eq:eqAP10}
\mathbf{J}=\frac{\partial \mathbf{h}}{\partial \boldsymbol{\Theta}}^H\frac{\partial \mathbf{h}}{\partial \boldsymbol{\Theta}}
\end{equation}
where
\begin{align}\label{eq:eqAP11}
\frac{\partial \mathbf{h}}{\partial \boldsymbol{\Theta}}&=\left[\frac{\partial \mathbf{h}}{\partial \mathfrak{R}(\boldsymbol{\beta})}\quad\frac{\partial \mathbf{h}}{\partial \mathfrak{I}(\boldsymbol{\beta})}\quad\frac{\partial \mathbf{h}}{\partial \boldsymbol{\mu}^{\mathrm{r}}}\quad\frac{\partial \mathbf{h}}{\partial \boldsymbol{\mu}^{\mathrm{t}}}\quad\frac{\partial \mathbf{h}}{\partial \boldsymbol{\nu}}\quad\frac{\partial \mathbf{h}}{\partial \boldsymbol{\eta}}\right]\nonumber\\
&=\left[\mathbf{A} \,\, j\mathbf{A} \,\, (\mathbf{D}_{\mathrm{r}}\diamond\mathbf{A}_{\mathrm{t}}\diamond
\mathbf{A}_{\mathrm{d}}\diamond\mathbf{A}_{\mathrm{f}})
\mathbf{X}\,\, (\mathbf{A}_{\mathrm{r}}\diamond\mathbf{D}_{\mathrm{t}}\diamond\mathbf{A}_{\mathrm{d}}\diamond\mathbf{A}_{\mathrm{f}})\mathbf{X}\right.\nonumber\\
&\qquad\left.(\mathbf{A}_{\mathrm{r}}\diamond\mathbf{A}_{\mathrm{t}}\diamond\mathbf{D}_{\mathrm{d}}\diamond\mathbf{A}_{\mathrm{f}})\mathbf{X}\quad
(\mathbf{A}_{\mathrm{r}}\diamond\mathbf{A}_{\mathrm{t}}\diamond\mathbf{A}_{\mathrm{d}}\diamond\mathbf{D}_{\mathrm{f}})\mathbf{X}\right]
\end{align}
With $\mathbf{G}_1$--$\mathbf{G}_5$ as given in \eqref{eqP11}--\eqref{eqP15}, \eqref{eq:eqAP11} can be simplified to
\begin{equation}
\frac{\partial \mathbf{h}}{\partial \boldsymbol{\Theta}}=\mathbf{G}_1\diamond\mathbf{G}_2\diamond\mathbf{G}_3\diamond\mathbf{G}_4\diamond\mathbf{G}_5
\end{equation}
and \eqref{eq:eqAP10} becomes
\begin{equation}\label{eq:eqAP23}
\mathbf{J}=(\mathbf{G}_1\diamond\mathbf{G}_2\diamond\mathbf{G}_3\diamond\mathbf{G}_4\diamond\mathbf{G}_5)^H(\mathbf{G}_1\diamond\mathbf{G}_2\diamond\mathbf{G}_3\diamond\mathbf{G}_4\diamond\mathbf{G}_5)
\end{equation}
Using the properties of the Khatri-Rao and Hardamad products, \eqref{eq:eqAP23} reduces to the form given in \eqref{eq:eq40}.


\ifCLASSOPTIONcaptionsoff
  \newpage
\fi



\bibliographystyle{IEEEtran}
\balance
\bibliography{sample1}

\begin{thebibliography}{10}
\providecommand{\url}[1]{#1}
\csname url@samestyle\endcsname
\providecommand{\newblock}{\relax}
\providecommand{\bibinfo}[2]{#2}
\providecommand{\BIBentrySTDinterwordspacing}{\spaceskip=0pt\relax}
\providecommand{\BIBentryALTinterwordstretchfactor}{4}
\providecommand{\BIBentryALTinterwordspacing}{\spaceskip=\fontdimen2\font plus
\BIBentryALTinterwordstretchfactor\fontdimen3\font minus
  \fontdimen4\font\relax}
\providecommand{\BIBforeignlanguage}[2]{{%
\expandafter\ifx\csname l@#1\endcsname\relax
\typeout{** WARNING: IEEEtran.bst: No hyphenation pattern has been}%
\typeout{** loaded for the language `#1'. Using the pattern for}%
\typeout{** the default language instead.}%
\else
\language=\csname l@#1\endcsname
\fi
#2}}
\providecommand{\BIBdecl}{\relax}
\BIBdecl

\bibitem{li2006}
G.~Li and G.~St{\"u}ber, \emph{{Orthogonal Frequency Division Multiplexing for
  Wireless Communications}}, ser. Signals and Communication Technology.\hskip
  1em plus 0.5em minus 0.4em\relax Springer Science+Business Media, 2006.

\bibitem{Dahlman2008}
E.~Dahlman, S.~Parkvall, J.~Skold, and P.~Beming, \emph{{3G Evolution, Second
  Edition: HSPA and LTE for Mobile Broadband}}, 2nd~ed.\hskip 1em plus 0.5em
  minus 0.4em\relax Academic Press, 2008.

\bibitem{Wimax1}
\emph{{Part 16: Air Interface for Broadband Wireless Access Systems}}, ser.
  IEEE Standard for local and metropolitan area networks.\hskip 1em plus 0.5em
  minus 0.4em\relax IEEE, {May} 2009.

\bibitem{Gray06}
D.~G. Edt, \emph{Mobile WiMAX – Part I: A Technical Overview and Performance
  Evaluation}, WiMAX Forum, 2006.

\bibitem{Xia04}
P.~Xia, S.~Member, S.~Zhou, and G.~B. Giannakis, ``{Adaptive MIMO-OFDM Based on
  Partial Channel State Information},'' \emph{IEEE Trans. Sig. Proc.}, vol.~52,
  pp. 202--213, 2004.

\bibitem{Kountouris2008}
M.~Kountouris, T.~S\"{a}lzer, and D.~Gesbert, ``{Scheduling for multiuser MIMO
  downlink channels with ranking-based feedback},'' \emph{EURASIP J. Adv.
  Signal Process}, vol. 2008, pp. 131:1--131:14, {Jan} 2008.

\bibitem{DBLP11}
S.~B. Lande, J.~B. Helonde, R.~Pande, and S.~S. Pathak, ``{Adaptive Subcarrier
  and Bit Allocation for Downlink OFDMA System with Proportional Fairness},''
  \emph{CoRR}, vol. abs/1111.2160, 2011.

\bibitem{Duel2000}
A.~Duel-Hallen, S.~Hu, and H.~Hallen, ``{Long Range Prediction of Fading
  Signals: Enabling Adaptive Transmission for Mobile Radio Channels},''
  \emph{IEEE Sig. Proc. Magazine}, vol.~17, pp. 62--75, 2000.

\bibitem{Rhee20082095}
D.~Rhee, H.~Hwang, Y.~Sang, and K.~Kim, ``{Multiuser adaptive transmission
  technique for time-varying frequency-selective fading channels},''
  \emph{Signal Processing}, vol.~88, no.~8, pp. 2095--2107, 2008.

\bibitem{Ekman2002}
T.~Ekman, ``{Prediction of Mobile Radio Channels --- Modeling and Design},''
  Ph.D. dissertation, Uppsala University, 2002.

\bibitem{Oien2004}
G.~E. Oien, H.~Holm, and K.~J. Hole, ``{Impact of channel prediction on
  adaptive coded modulation performance in Rayleigh fading},'' \emph{IEEE
  Trans. on Vehicular Technology}, vol.~53, pp. 758--769, 2004.

\bibitem{Anderson99}
J.~Andersen, J.~Jensen, S.~Jensen, and F.~Frederisen, ``{Prediction of future
  fading based on past measurements},'' in \emph{IEEE Vehicular Technology
  Conference}, vol.~1, 1999, pp. 151--155.

\bibitem{Wong08}
I.~C. Wong and B.~L. Evans, ``{Sinusoidal Modeling and Adaptive Channel
  Prediction in Mobile OFDM Systems},'' \emph{IEEE Trans. on Sig. Proc.},
  vol.~56, no.~4, pp. 1601--1615, 2008.

\bibitem{Vaughan}
R.~Vaughan, P.~Teal, and R.~Raich, ``{Short-term mobile channel prediction
  using discrete scatterer propagation model and subspace signal processing
  algorithms},'' in \emph{IEEE Vehicular Technology Conference}, vol.~2, 2000,
  pp. 751--758.

\bibitem{Liu2006}
J.~Liu and X.~Liu, ``{Time-varying channel identification and prediction in
  OFDM systems using 2-D frequency estimation},'' in \emph{Proceedings of the
  2006 IEEE Conference on Military Communications}, ser. MILCOM'06.\hskip 1em
  plus 0.5em minus 0.4em\relax Piscataway, NJ, USA: IEEE Press, 2006, pp.
  2777--2783.

\bibitem{Yang2001}
B.~Yang, K.~B. Letaief, R.~S. Cheng, and Z.~Cao, ``{Channel estimation for OFDM
  transmission in multipath fading channels based on parametric channel
  modeling},'' \emph{IEEE Trans. on Communications}, vol.~49, no.~3, pp.
  467--479, aug 2002.

\bibitem{paul2001}
P.~Teal and R.~Vaughan, ``{Simulation and performance bounds for real-time
  prediction of the mobile multipath channel},'' in \emph{{IEEE Workshop on
  Statistical Signal Processing Proceedings}}, 2001, pp. 548--551.

\bibitem{Arredondo02}
A.~Arredondo, K.~R. Dandekar, and G.~Xu, ``{Vector channel modeling and
  prediction for the improvement of downlink received power},'' \emph{IEEE
  Trans. on Comm.}, vol.~50, no.~7, pp. 1121--1129, 2002.

\bibitem{Svantesson2006}
T.~Svantesson and A.~Swindlehurst, ``{A performance bound for prediction of
  MIMO channels},'' \emph{IEEE Trans. on Sig. Proc.}, vol.~54, no.~2, pp.
  520--529, {Oct} 2006.

\bibitem{Larsen2008}
M.~Larsen, L.~Swindlehurst, and T.~Svantesson, ``{A Performance Bound for
  MIMO-OFDM Channel Estimation and Prediction},'' in \emph{Proc. Fifth IEEE
  Sensor Array and Multichannel Signal Processing Workshop}, 2008, pp.
  141--145.

\bibitem{kenta08}
K.~Okino, T.~Nakayama, S.~Joko, Y.~Kusano, and S.~Kimura, ``{Direction based
  beamspace MIMO channel prediction with ray cancelling},'' in \emph{Proc. IEEE
  PIMRC}, 2008, pp. 1--5.

\bibitem{ChangCodebook11}
J.~Chang, I.-T. Lu, and Y.~Li, ``{Adaptive Codebook Based Channel Prediction
  and Interpolation for Multiuser MIMO-OFDM Systems},'' in \emph{ICC}, 2011,
  pp. 1--5.

\bibitem{Inoue2009}
T.~Inoue and R.~W. Heath, Jr., ``{Grassmannian Predictive Coding for delayed
  limited feedback MIMO systems},'' in \emph{Proceedings of the 47th annual
  Allerton conference on Communication, control, and computing}, ser.
  Allerton'09.\hskip 1em plus 0.5em minus 0.4em\relax Piscataway, NJ, USA: IEEE
  Press, 2009, pp. 783--788.

\bibitem{Godana2011}
B.~Godana and T.~Ekman, ``{Linear prediction of time-varying MIMO systems using
  Givens rotations},'' in \emph{IEEE Workshop on Signal Processing Advances in
  Wireless Communications}, 2011, pp. 371--375.

\bibitem{Vanderpypen}
J.~Vanderpypen and L.~Schumacher, ``{MIMO Channel Prediction using ESPRIT based
  Techniques},'' in \emph{{Proc. IEEE PIMRC}}.

\bibitem{Larsen2009}
M.~D. Larsen, A.~L. Swindlehurst, and T.~Svantesson, ``{Performance bounds for
  MIMO-OFDM channel estimation},'' \emph{IEEE Trans. on Sig. Proc.}, vol.~57,
  no.~5, pp. 1901--1916, {May} 2009.

\bibitem{Roy2}
R.~Roy and T.~Kailath, ``Estimation of signal parameters via rotational
  invariance techniques,'' \emph{IEEE Trans on Acoustics, Speech, Signal
  Processing}, vol.~37, pp. 984--995, Jul 1989.

\bibitem{WINNER2}
``{IST-WINNER II Deliverable 1.1.2 v.1.2, WINNER II channel models},''
  {IST-WINNER2}, Tech. Rep., 2007.

\bibitem{Wax85}
M.~Wax and T.~Kailath, ``{Detection of signals by information theoretic
  criteria},'' \emph{IEEE Trans. on Acoustics, Speech and Signal Processing},
  vol.~33, no.~2, pp. 387--392, 1985.

\bibitem{Kay}
S.~M. Kay, \emph{{Modern Spectral Estimation: Theory and Application}}.\hskip
  1em plus 0.5em minus 0.4em\relax Prentice Hall.

\bibitem{Huang_mmsemdl}
L.~Huang, T.~Long, E.~Mao, H.~C. So, and S.~Member, ``{MMSE-Based MDL Method
  for Accurate Source Number Estimation},'' \emph{IEEE Signal Processing
  Letters}, vol.~16, no.~9, pp. 798--801, 2009.

\bibitem{Kikuma}
N.Kikuma, H.Kikuchi, and N.Inagaki, ``{Pairing of Estimates Using Mean
  Eigenvalue Decomposition in Multi-Dimensional Unitary ESPRIT},'' \emph{IEICE
  Trans}, vol. J82-B, no.~11, pp. 2202--2207, {Nov} 1999.

\bibitem{Kay1}
S.~M. Kay, \emph{{Fundamentals of statistical signal processing: estimation
  theory}}.\hskip 1em plus 0.5em minus 0.4em\relax Upper Saddle River, NJ, USA:
  Prentice-Hall, Inc., 1993.

\bibitem{WINNER}
P.~Kyösti, J.~Meinilä, L.~Hentilä, X.~Zhao, T.~Jämsä, C.~Schneider,
  M.~Narandzić, M.~Milojević, A.~Hong, J.~Ylitalo, V.-M. Holappa,
  M.~Alatossava, R.~Bultitude, Y.~de~Jong, and T.~Rautiainen, ``{WINNER II
  Channel Models},'' Tech. Rep., Sep 2007.

\end{thebibliography}

%








\end{document}